\documentclass[%
 reprint,
 amsmath,amssymb,
 aps,
]{revtex4-1}

\usepackage[english]{babel}
\usepackage[utf8]{inputenc}
\usepackage[caption=false]{subfig}
\usepackage{graphicx}
\usepackage{dcolumn}
\usepackage{bm}
\usepackage{color}
\usepackage{latexsym}
\usepackage{amsmath}
\usepackage{amssymb}
\usepackage{amsfonts}
\usepackage[space-before-unit,range-units = repeat]{siunitx}

\newcommand {\txt}[1]{\text{#1}}
\DeclareSIUnit\atmosphere{atm}

\begin{document}

\preprint{APS/123-QED}

  \title{Thin-sheet creation and threshold pressures in drop splashing}

\author{Andrzej Latka}
 \affiliation{The James Franck Institute and Department of Physics,The University of Chicago, Chicago, Illinois 60637, USA}

\date{\today}

\begin{abstract}

A liquid drop impacting a smooth solid substrate splashes by emitting a thin liquid sheet from near the contact line of the spreading liquid. This sheet is lifted from the substrate and ultimately breaks apart. Surprisingly, the splash is caused by the ambient gas, whose properties dictate when and if the sheet is created. Here I focus on two aspects of this process. Using high-speed imaging I find that the time of thin-sheet creation displays a different quantitative dependence on air pressure if the sheet is created during the early stages of spreading, rather than when the liquid has already spread to a large radius. This result sheds light on previously observed impact velocity regimes. Additionally, by measuring impacts of drops on surfaces comprised of both rough and smooth regions, I identify a new threshold velocity that limits the times at which the thin sheet can be created. This velocity determines the threshold pressure below which splashing is suppressed.  

\end{abstract}

\maketitle


\section{\label{sec:introduction}Introduction}

The splash of a liquid drop on a dry smooth surface is caused by the ambient air \cite{Xu2005}. Experiments have found that drops splash only above a certain gas pressure in a variety of systems \cite{Xu2005, Xu2007, Xu2007a, Tsai2010, Latka2012}, and have provided insight into the liquid and air dynamics during splashing \cite{Driscoll2010, Kolinski2012,bischofberger2016airflows,liu2015kelvin}. Recently, the air effect was reproduced numerically \cite{boelens2016observation,guo2016investigation}. Nevertheless, proposed theories \cite{Mandre2009, Riboux2014} have been unsuccessful in describing the experimental data. The mechanism by which air causes splashing is still unclear. 

\begin{figure}
\includegraphics[width=1\columnwidth]{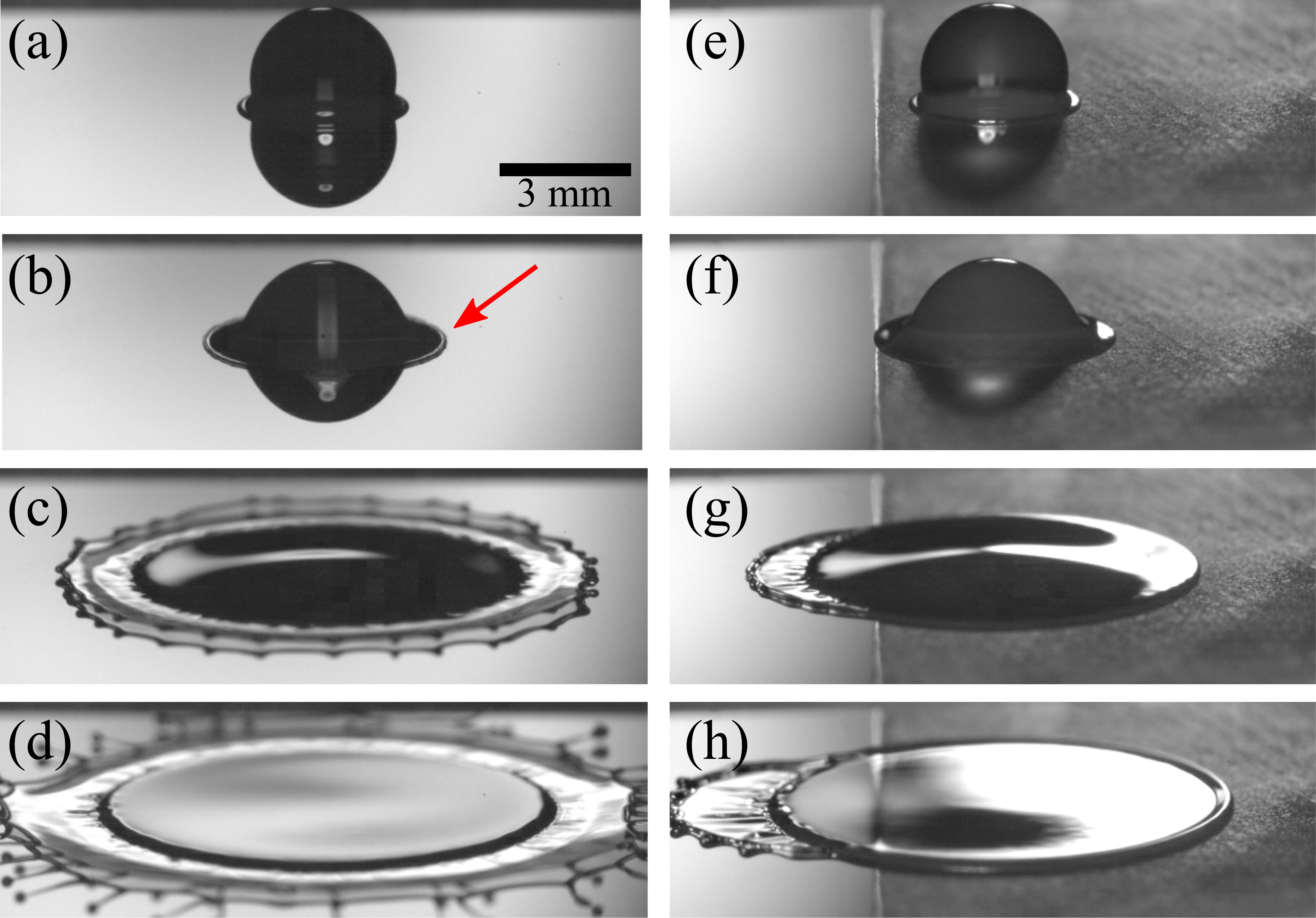}
\caption[]{\label{fig:images}
Successive images of a $9.4 \si{\milli Pa.s}$ silicone oil drop of diameter $D=3.3 \si{\milli m}$ impacting a glass slide at $V=3.4 \si{\m.s^{-1}}$ ($\txt{Re}=1100$, $\txt{We}=1600$) at atmospheric pressure. Images (a)-(d) show the drop splashing on a smooth glass slide at times $t = 0.18, 0.33, 1.2$ and $2.3 \si{\milli s}$. The red arrow in (b) points to the newly-created thin sheet that grows in subsequent images. Images (e)-(f) show the corresponding frames of a drop impacting a slide that is comprised of a rough (right, dark) and a smooth (left, bright) region. Thin-sheet formation and splashing take place only in the smooth region. 
}
\end{figure}

The splash of the liquid drop occurs in multiple stages. Shortly before impact, the bottom surface of the drop is deformed by the rising gas pressure in the decreasing gap between the liquid and solid \cite{Mandre2009,mandre2012mechanism}. Consequently, when the liquid finally makes contact with the substrate \cite{Kolinski2012}, the air directly beneath the drop is trapped in a small bubble \cite{Thoroddsen2010} and does not further influence the splashing process \cite{Driscoll2011}. Next, the liquid spreads radially outward in the form of a liquid sheet that remains in direct contact with the substrate, as shown in Fig.\ \ref{fig:images}a \cite{Driscoll2010,Driscoll2011,Stevens2015}.  At a time $t_\txt{sheet}$ after impact, the advancing liquid abruptly begins to move over a layer of air approximately several microns thick \cite{Driscoll2011}. The creation of this air gap leads to the ejection of a thin sheet of liquid \cite{Driscoll2010, Driscoll2011}, as shown in Fig.\ \ref{fig:images}b. It is the subsequent breakup of this thin sheet that finally results in a splash, as shown in Fig.\ \ref{fig:images}c-d. If the pressure is decreased below a threshold $P_\txt{sheet}$, the thin sheet is never created and the splash is suppressed. The liquid simply continues to spread radially outward in contact with the substrate until it comes to rest. Clearly, the formation of the splash hinges upon the air-induced creation of the thin sheet. The majority of present theories of splashing do not take this mechanism into account. By considering the role of thin-sheet creation in splashing, I am able to shed light on two outstanding questions: why two impact velocity regimes of threshold pressure exist \cite{Driscoll2010}, and why splashing is suppressed at low pressures. 

In the first half of this work, I focus on how the process of thin-sheet creation changes with the pressure of the ambient gas. I find that this air dependence is markedly different at the early stages of impact, when the drop has not yet spread significantly on the surface, than at later stages, when the radius of the region wetted by the liquid is much larger than the original drop’s radius. Distinct dependence of the thin-sheet formation onset on pressure for the "small-radius" and "large-radius" sheets explains the existence of a high and low impact velocity regime. 

In the second half of this paper, I focus on how $P_\txt{sheet}$ is determined in the high impact velocity regime. It had previously been shown \cite{Latka2012} that thin-sheet formation can be suppressed if a drop impacts a rough surface. I use this effect to delay thin-sheet creation in a spreading drop, by letting it fall on a substrate comprised of a rough and a smooth region. The right column of Fig.\ \ref{fig:images} shows that a thin sheet is created only on the smooth side of the surface. By changing the point of impact, I can control the velocity of the contact line when it first reaches the smooth region and find that the thin sheet can only be created, when the contact line is moving faster than a threshold velocity $u_\txt{stop}$. The velocity $u_\txt{stop}$ is independent of the ambient air. Together, the air-dependent time of thin-sheet creation and the air-independent threshold velocity, below which thin-sheet creation is suppressed, form a pair of necessary and sufficient conditions for thin-sheet creation. The threshold pressure is the pressure below which both conditions cannot be simultaneously satisfied.  

\section{\label{sec:experimental}Experimental details}

The experiments were conducted with a variety of liquids. Ethanol and silicone oils (PDMS, Clearco Products) were used to vary the drop viscosity from $\mu = 1.2$ to $48 \si{\milli Pa.s}$, while keeping the surface tension approximately constant between $\sigma = 18.7$ and $21.6 \si{\milli \N. \m^{-1}}$. Solutions of water/glycerol or ethanol/water/sucrose were used to study the effect of increasing the surface tension to $66 \si{\milli \N. \m^{-1}}$ or $55 \si{\milli \N. \m^{-1}}$, respectively. The effect of increasing density to $\rho=1900 \si{\kg. \m^{-3}}$ was measured with Fluorinert (3M Fluorinert Electronic Liquid). 

Drops with diameters $D$ ranging from $1.7$ to $3.7 \si{\milli \m}$ were produced using a syringe pump (Razel Scientific, Model R99-E) and nozzles of varying sizes. The drops were then released from a nozzle above a substrate. The height at which the nozzle is positioned above the surface sets the impact velocity $V$, which was varied between $1.5 \si{\m.\s^{-1}}$ and $4.1 \si{\m.\s^{-1}}$. These parameters can be summarized by the Reynolds number, $\txt{Re} = \frac{\rho D V}{\mu}$, which gives the ratio of inertial to viscous forces, and the Weber number, $\txt{We}=\frac{\rho D V^2}{\sigma}$, which gives the ratio of inertial to surface tension forces. Here we consider the regime $47<\txt{Re}<7700$ and $110<\txt{We}<2500$.

The drops impacted glass slides (Fisherbrand Microscope Slides) that were either uniformly smooth, as seen in the left column of Fig.\ \ref{fig:images}, or were divided into one smooth and one rough region, as seen in the right column of that figure. The patterned slides were created by etching half of a slide with ammonium bifluoride for 8 minutes (Armour Etch). The time of etching was chosen to produce a root-mean-square roughness $R_\txt{rms} = 1.9 \si{\micro \m}$, sufficient to prevent thin-sheet creation for the liquids used \cite{Latka2012}. The surface was characterized with atomic force microscopy (Asylum MFP-3D AFM) on $75 \si{\micro \m}$ square patches. 

The experiments were conducted in a vacuum chamber, with gas pressures that could be varied between $P=5 \si{\kilo Pa}$ and $101 \si{\kilo Pa}$. Three different gases were used to measure the effect of gas molecular weight $M_\txt{w}$ and viscosity $\mu_\txt{g}$: air ($29 \si{g.mol^{-1}}$, $18.6 \si{\micro Pa.s}$), helium ($4.0 \si{g.mol^{-1}}$, $20.0 \si{\micro Pa.s}$), and neon ($20.2 \si{g.mol^{-1}}$, $32.1 \si{\micro Pa.s}$).

The onset of thin-sheet creation is marked by the appearance of an air gap between the spreading liquid and the substrate, therefore the accurate detection of this air gap ensures a precise measurement of $t_\txt{sheet}$. This accuracy was provided by ultra-fast interference imaging, which measures the interference between the light reflected from the bottom surface of the spreading liquid and the top surface of the substrate \cite{Driscoll2011}. When the liquid is in contact with the substrate, little light is reflected back to the camera, since the index of refraction of glass slides ($n_g = 1.5$) and silicone oil ($n_l = 1.4$) are similar. However, when the liquid is separated from the substrate by a layer of air, an interference pattern is created, as seen in Fig.\ \ref{fig:interference}. Several factors determine the resolution of this method. The camera resolution (Vision Research v12, v1610, v2512) was $17.6$-$24.3 \si{\micro m}$ per pixel and frame rates up to $300000$ fps with $0.25 \si{\micro s}$ exposure time were used. The thinnest air gap we can reliably detect, approximately $30 \si{\nano m}$, is set by the wavelength of the light source (ThorLabs LED, $\lambda = 625 \si{\nano m}$), the sensitivity of the camera, and the exposure time. The thickest gap that can be measured is set by the coherence length of the light source, approximately $10 \si{\micro m}$. This is not a limiting factor, since the goal of these measurements was to identify the air gap as soon as it was created. Finally, interference fringes will not be resolved if the slope of the thin sheet's bottom surface is large enough ($>\ang{0.3}$) that multiple fringes are recorded at a single camera pixel. The slope of the thin sheet increases with pressure and with decreasing liquid viscosity, therefore drops with $\mu < 2 \si{\milli Pa.s}$ are difficult to image at large pressures. For a detailed explanation of ultra-fast interference imaging, a diagram of the experimental setup, and calculations of the reflected light intensity, see \cite{Driscoll2011}. 

The interference measurements were complemented with recordings of drop impacts from the side, which allowed for direct identification of the thin sheet, as well as for observing the spreading of the drop on rough surfaces, whose geometry makes interference imaging impossible.

\section{\label{sec:early_and_late_splashes}Time of thin sheet creation}

\begin{figure}
\includegraphics[width=\columnwidth]{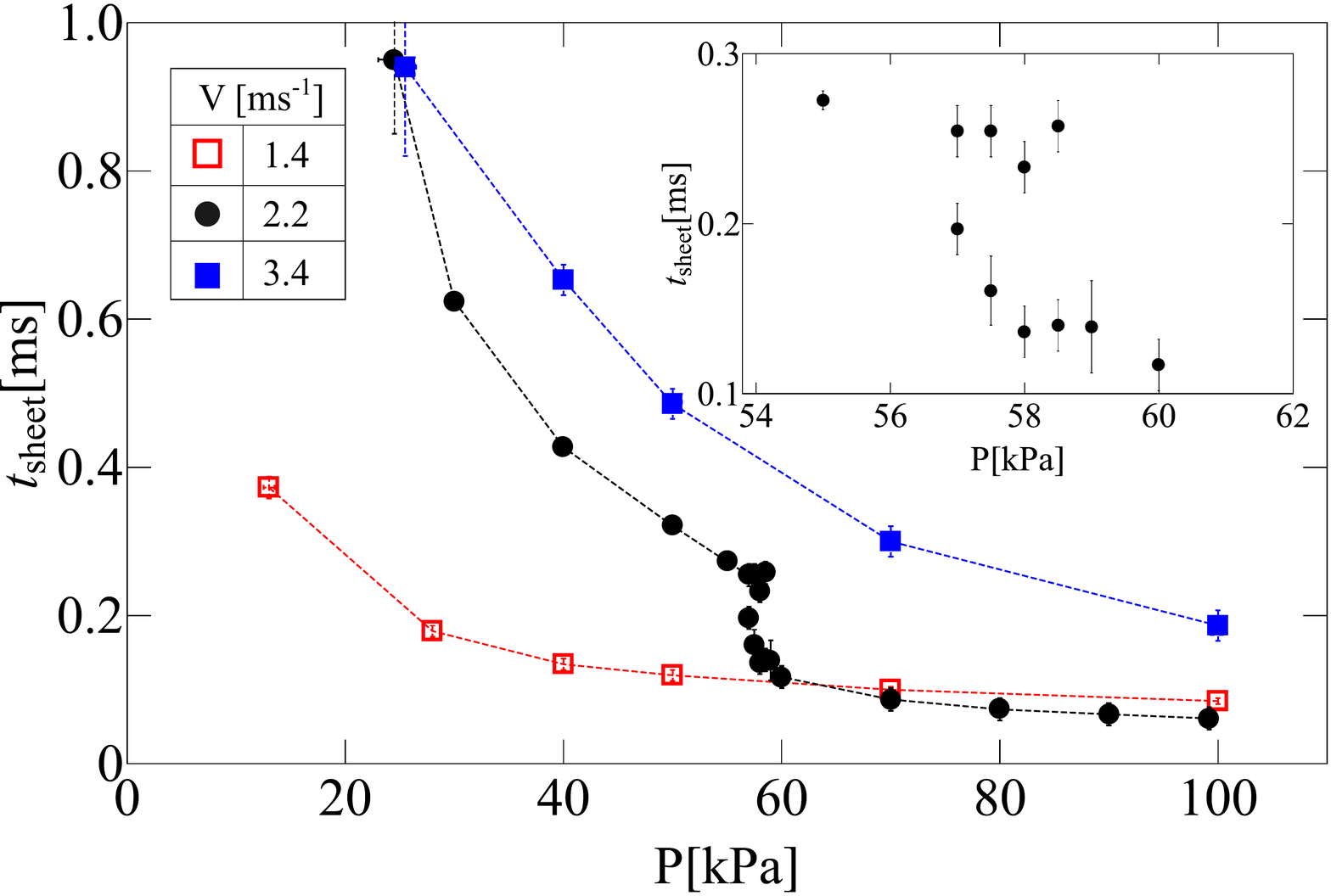}
\caption[]{\label{fig:tSheet_237}
Sheet creation time vs. pressure for $9.4 \si{\milli Pa.s}$ silicone oil drops of radius $R=1.65 \si{\m.s^{-1}}$ impacting a glass slide with velocity $1.4 \si{\m.s^{-1}}$ (\includegraphics[height=6pt]{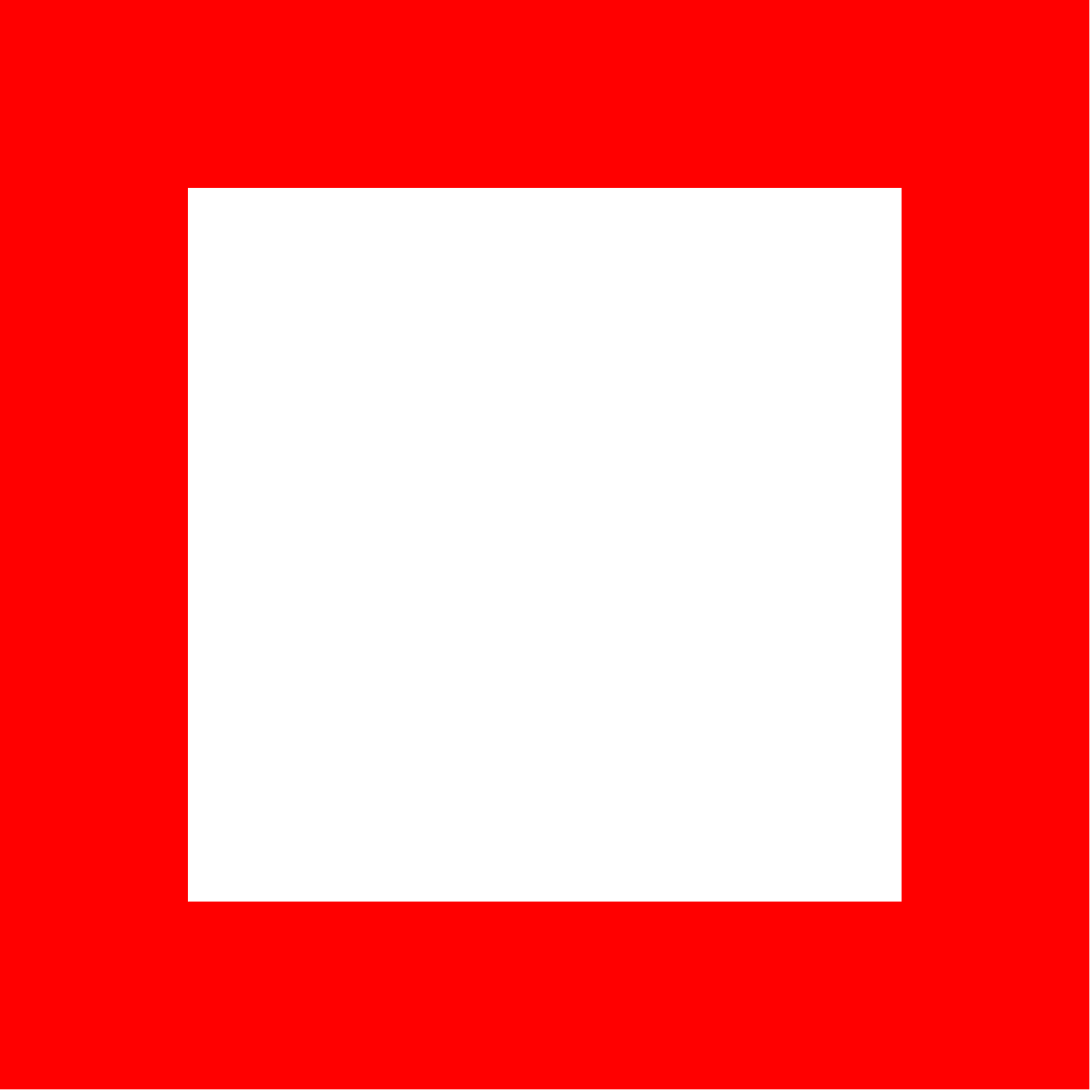}), $2.2 \si{\m.s^{-1}}$ (\includegraphics[height=6pt]{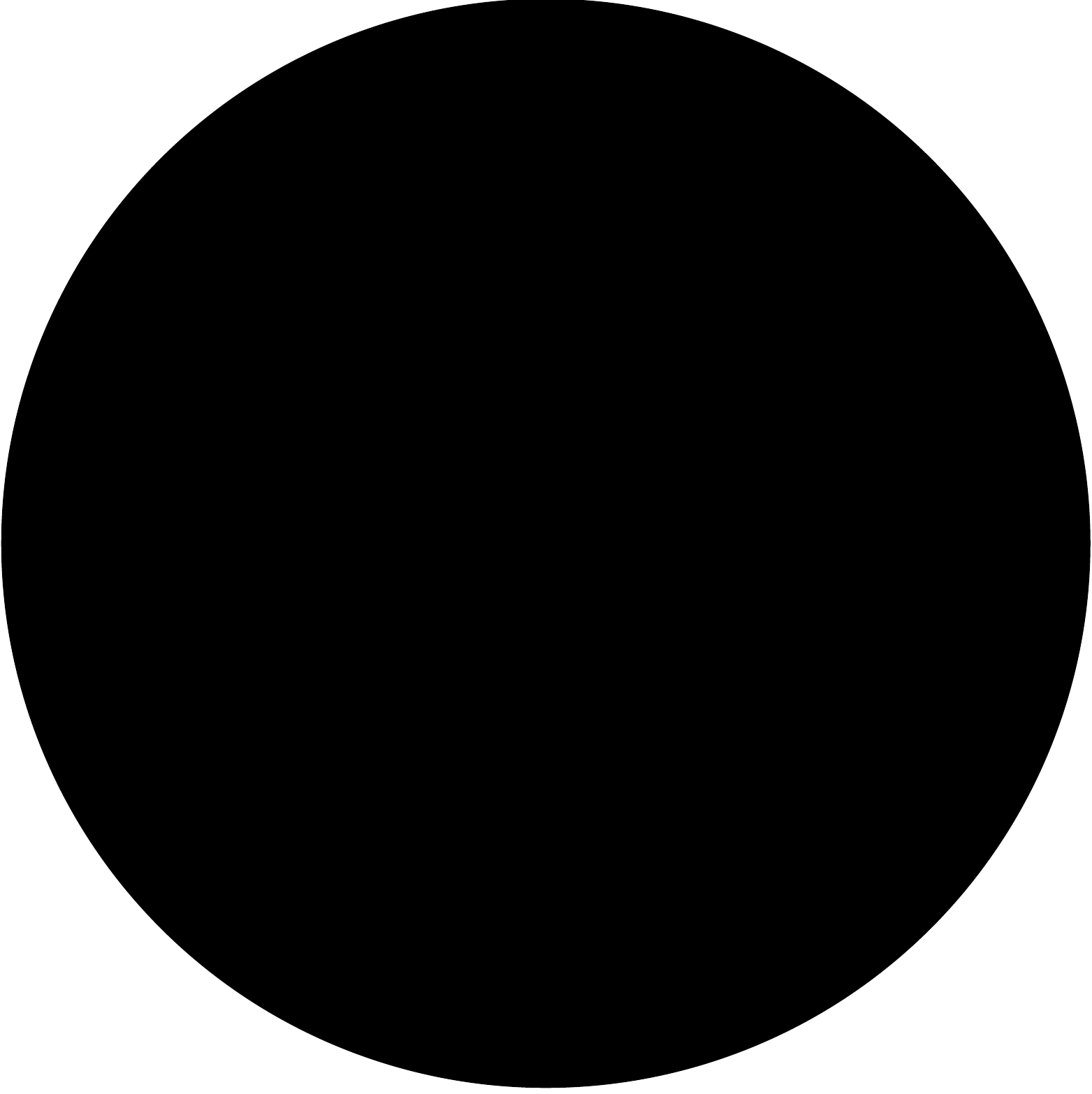}), and $3.4 \si{\m.s^{-1}}$ (\includegraphics[height=6pt]{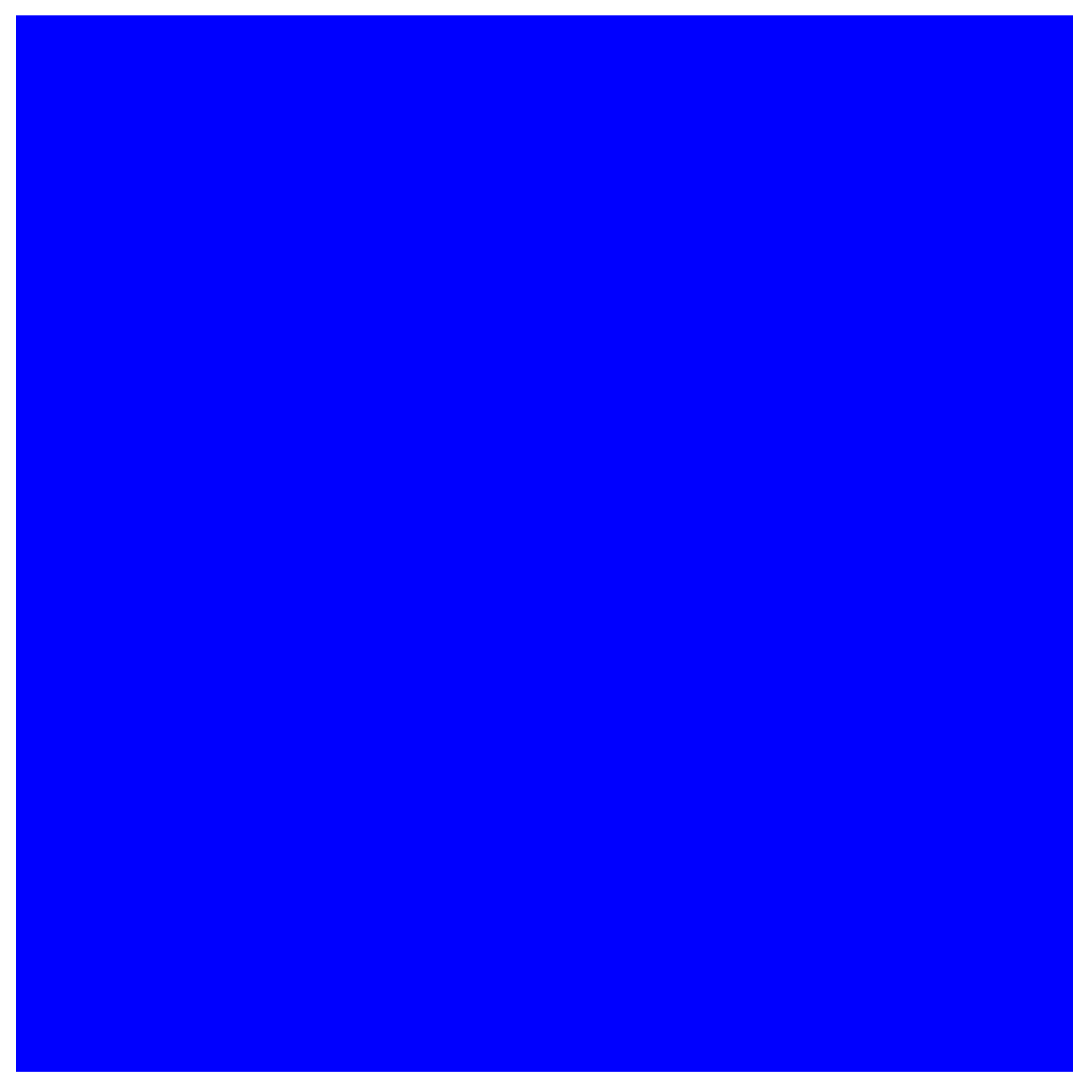}). The increase in $t_\txt{sheet}$ with decreasing pressure is smooth for both the fastest and the slowest drops. In contrast, a transition in $t_\txt{sheet}$ is seen at $P = 58 \pm 1 \si{\kilo Pa}$ for the intermediate velocity drop. Above the transition the sheet creation time slowly increases from $0.06 \si{\milli s}$ to $0.12 \si{\milli s}$ as pressure is reduced from $100 \si{\kilo Pa}$ to $60 \si{\kilo Pa}$. As the pressure is further reduced to $57 \si{\kilo Pa}$, $t_\txt{sheet}$ more than doubles to $0.27 \si{\milli s}$. The distribution of $t_\txt{sheet}$ is bimodal in this region, with sheet creation times clustered around either $0.12 \si{\milli s}$ or $0.27 \si{\milli s}$. As pressure is reduced further, $t_\txt{sheet}$ increases smoothly. The inset shows the transition region in detail. Lines are guides to the eye.  
}
\end{figure}

\begin{figure*}
\includegraphics[width=0.85\textwidth]{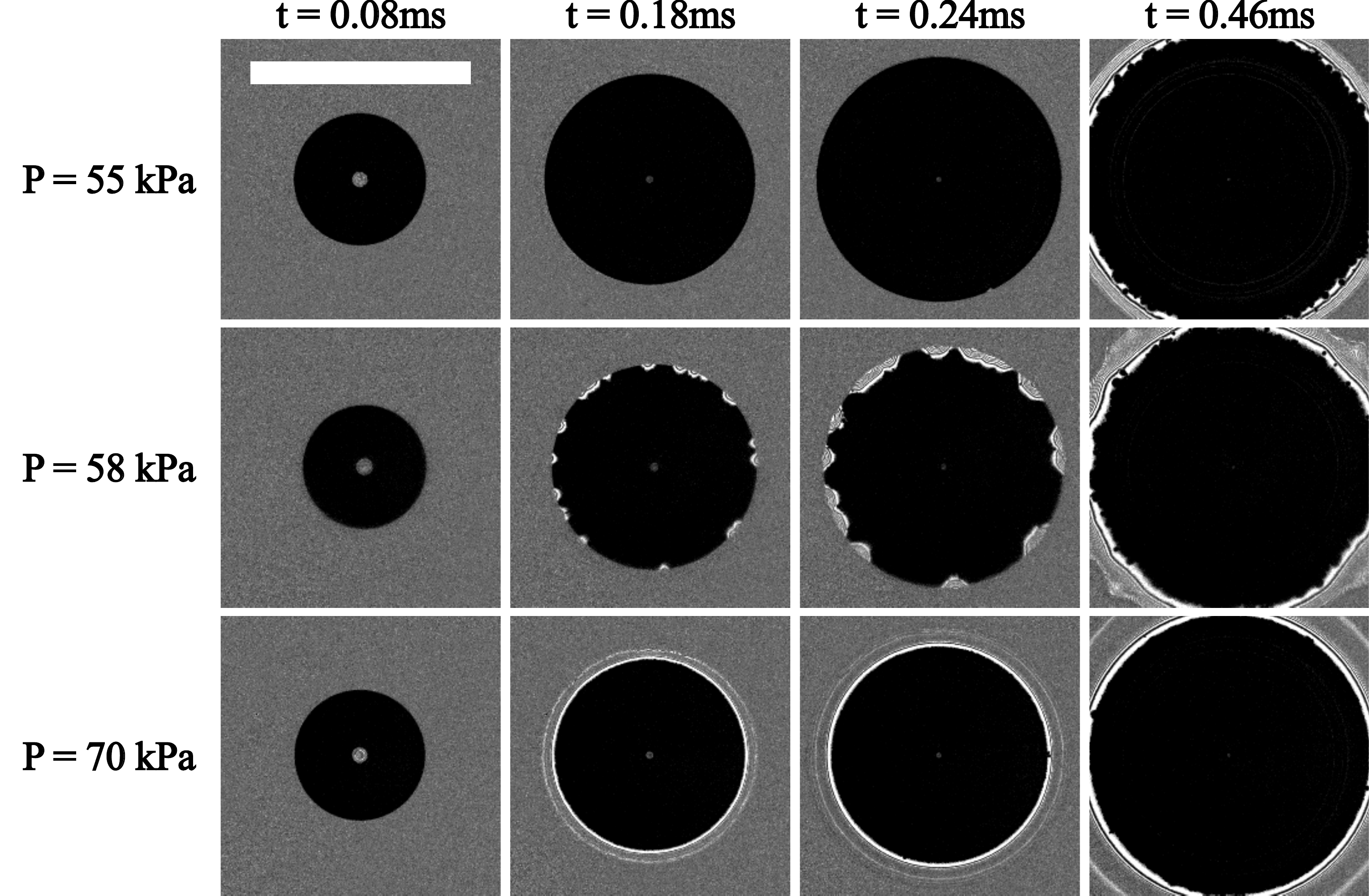}
\caption[]{\label{fig:interference}
Interference images of $9.4 \si{\milli Pa.s}$ silicone oil drops of radius $R=1.65 \si{\milli m}$ impacting a glass slide with velocity of $2.2 \si{\m.s^{-1}}$ as seen from below. The images were taken at $t = 0.08$, $0.18$, $0.24$, and $0.46 \si{\milli s}$ after impact and at three different pressures, which were chosen to illustrate the transition in Fig.\ \ref{fig:tSheet_237}: $P = 55 \si{\kilo Pa}$ below the transition pressure, $58 \si{\kilo Pa}$ in the transition region, and $70 \si{\kilo Pa}$ above the transition pressure. At the earliest time the image of the spreading liquid is black, indicating that it is in contact with the glass. A trapped air bubble can be seen trapped at the center of each impact \cite{Thoroddsen2010}. At $t = 0.18 \si{\milli s}$ an interference pattern is seen at the edge of the $70 \si{\kilo Pa}$ drop, indicating that the liquid is now spreading over a thin air gap and that a thin sheet has been created. The air gap is also seen in parts of the liquid edge at $58 \si{\kilo Pa}$. At $t = 0.24 \si{\milli s}$ the interference patterns grow for both the $58 \si{\kilo Pa}$ and the $70 \si{\kilo Pa}$ drops. In the $58 \si{\kilo Pa}$ images, the regions of the spreading drop that had not already formed a thin sheet remain on the substrate and do not form a thin sheet until a later time. At $t = 0.46 \si{\milli s}$ an air gap has finally developed for the $55 \si{\kilo Pa}$ drop, as well as the remaining regions of the $58 \si{\kilo Pa}$ case. Close inspection reveals that the liquid locally bridges the air gap near the contact line, as described in \cite{Driscoll2010}. The white scale bar is equivalent to $3.3 \si{\milli \m}$, the diameter of the original drop.
}
\end{figure*}

The thin-sheet creation time depends on a number of parameters, in particular on the ambient gas pressure \cite{Driscoll2010}. Figure \ref{fig:tSheet_237} shows that at $P=100 \si{\kilo Pa}$ a $9.4 \si{\milli Pa.s}$ silicone oil drop of radius $R=1.65 \si{\m.s^{-1}}$ that impacts a glass slide at $3.4 \si{\m.s^{-1}}$ creates a thin sheet at $t_\txt{sheet} = 0.019 \si{\milli s}$. If the pressure is decreased, $t_\txt{sheet}$ will smoothly increase, until below a pressure of $P_\txt{sheet}=25.5 \si{\kilo Pa}$ sheet creation is suppressed completely, so that no splashing can occur below this pressure. For a lower impact velocity of $1.4 \si{\m.s^{-1}}$ sheet creation occurs much earlier. Nevertheless, as pressure is decreased, $t_\txt{sheet}$ gradually increases for this $V$ as well. Both cases are consistent with \cite{Driscoll2010}.

A different behavior with decreasing pressure is observed for an intermediate impact velocity of $2.2 \si{\m.s^{-1}}$. As the pressure is lowered below atmospheric pressure, the sheet initially is created at times similar to the $V = 1.4 \si{\m.s^{-1}}$ case. However, as the pressure is decreased below $P = 58 \pm 1 \si{\kilo Pa}$, $t_\txt{sheet}$ rapidly increases and begins following a new trend, more similar to the $V = 3.4 \si{\m.s^{-1}}$ case. The inset of Fig.\ \ref{fig:tSheet_237} shows in detail the transition of $t_\txt{sheet} \left( P \right)$ between  the two trends. At both high and low pressures, the recorded values of $t_\txt{sheet}$ were distributed around a single value. Between $57 \si{\kilo Pa}$ and $59 \si{\kilo Pa}$, however, the measured distribution of $t_\txt{sheet}$ is bimodal, with drops creating sheets either at $t_\txt{sheet} \approx 0.25 \si{\milli \s}$, following the low pressure trend, or at $t_\txt{sheet} \approx 0.14 \si{\milli \s}$, consistent with sheet creation above the transition.  

Interference images shown in Fig.\ \ref{fig:interference} reveal that this transition can be observed during a single drop impact. The top row of images shows the spreading of the $V = 2.2 \si{\m.s^{-1}}$ drop at $P = 55 \si{\kilo Pa}$. Until $t_\txt{sheet} \left( 55 \si{\kilo Pa}\right) = 0.27 \si{\milli s}$, the liquid spreads smoothly on the substrate. Subsequently, a thin sheet is created, as evidenced by the interference pattern visible at $t = 0.46 \si{\milli s}$. The bottom row shows images of a drop above the transition, at $70 \si{\kilo Pa}$. Here, thin sheet creation occurs much earlier at $t_\txt{sheet} \left( 70 \si{\kilo Pa}\right) = 0.09 \si{\milli s}$. The middle row represents a drop at the transition pressure. The second interference image shows the air gap that was created at time $t = 0.14 \si{\milli s}$. Careful inspection reveals that contrary to what was seen at higher and lower pressures, the air gap is present only at certain points along the advancing contact line. At those points, the air gap continues to grow with time. The remaining points of the contact line are following the low-pressure trend: no air gap is present until $t = 0.24 \si{\milli s}$ and only then does this region begin to create a thin sheet.

\begin{figure}
\includegraphics[width=\columnwidth]{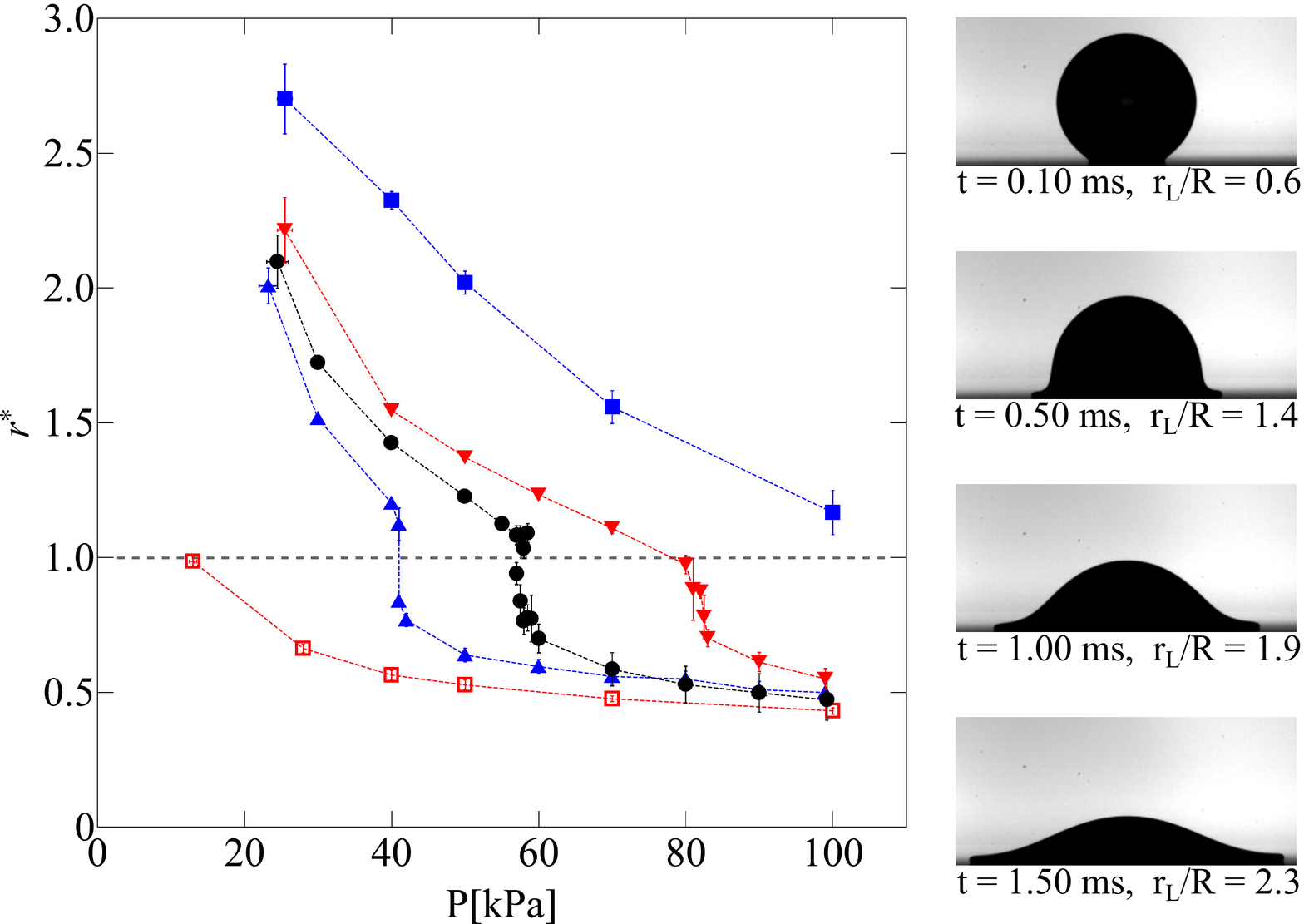}
\caption[]{\label{fig:rSheet}
The radius of region wetted by the drop at $t_\txt{sheet}$ rescaled by drop radius, $r^*$ vs. pressure for $9.4 \si{\milli Pa.s}$ silicone oil drops of radius $R=1.65 \si{\m.s^{-1}}$ impacting a glass slide with velocity: $1.4 \si{\m.s^{-1}}$ (\includegraphics[height=6pt]{EmptyRedSquare.pdf}), $1.9 \si{\m.s^{-1}}$ (\includegraphics[height=6pt]{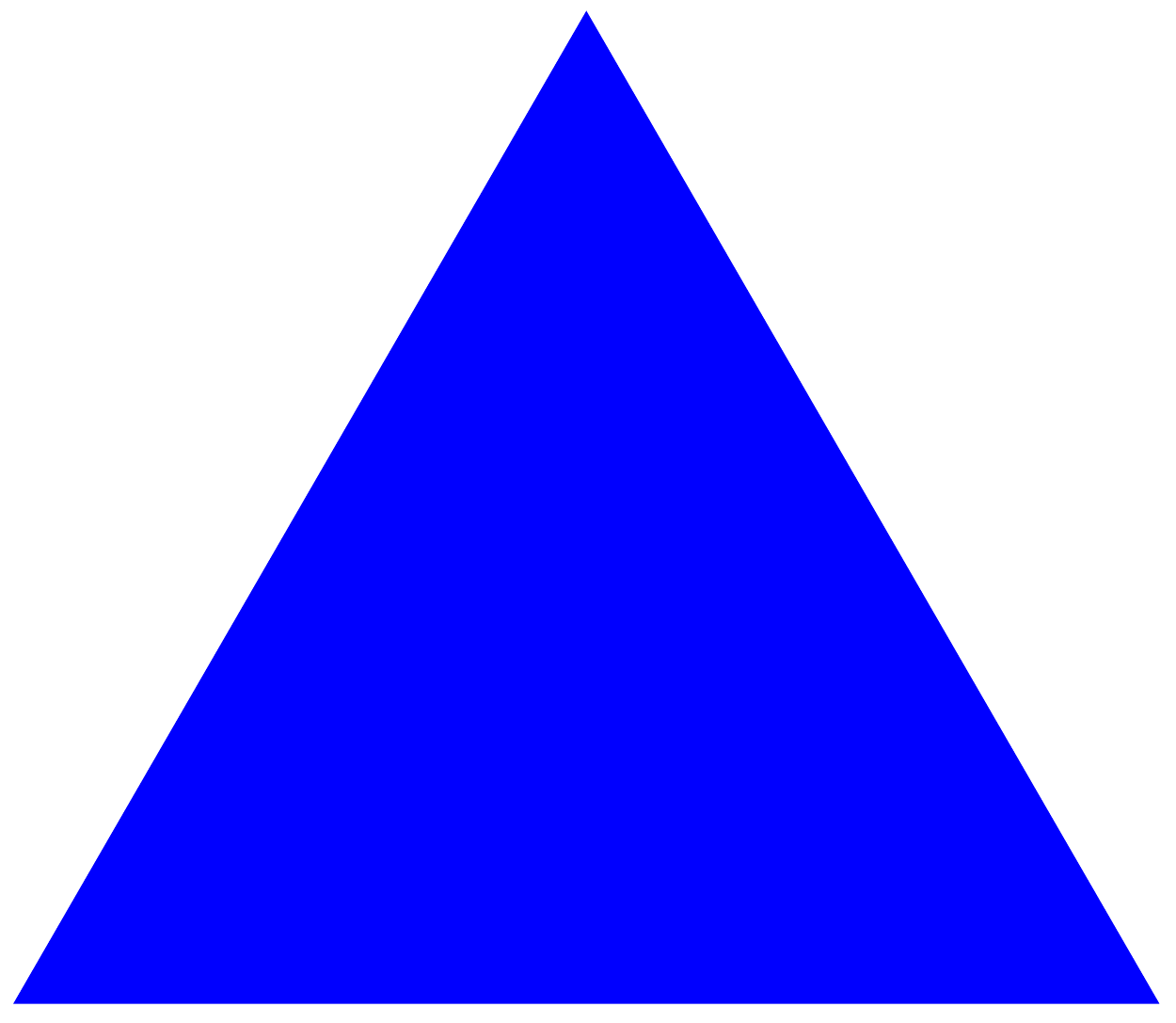}), $2.2 \si{\m.s^{-1}}$ (\includegraphics[height=6pt]{FilledBlackCircle.pdf}), $2.5 \si{\m.s^{-1}}$ (\includegraphics[height=6pt]{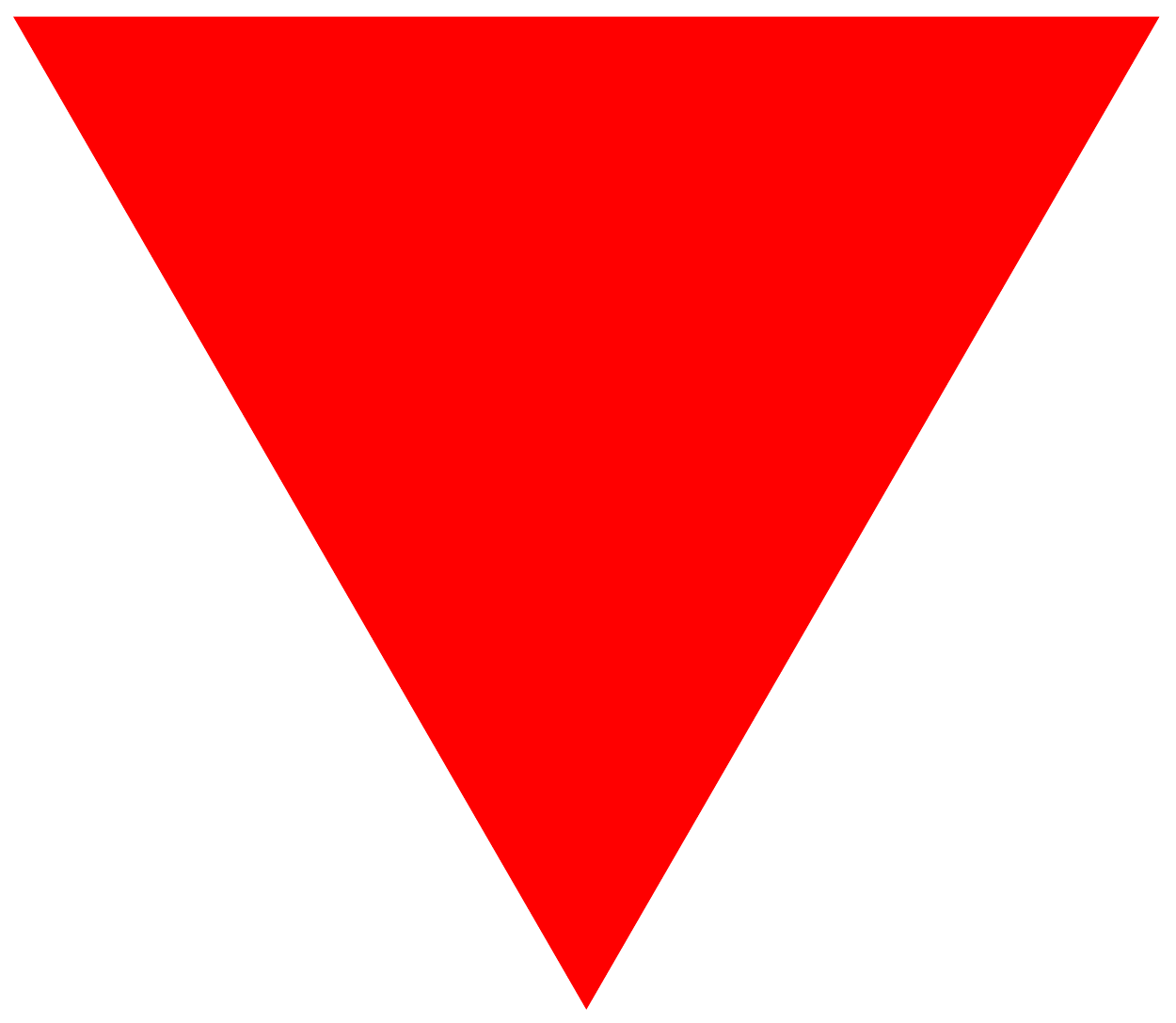}), and $3.4 \si{\m.s^{-1}}$ (\includegraphics[height=6pt]{FilledBlueSquare.pdf}). The transition occurs at $r^* \approx 1$. Lines are guides to the eye. The images show the difference in the shape of a $V = 1.8 \si{\m.s^{-1}}$ drop at times between $0.1 \si{\milli s}$ and $1.5 \si{\milli s}$, as seen from the side.
}
\end{figure}

Figure \ref{fig:rSheet} sheds light on this unusual behavior, by comparing how far the liquid has spread, $r_\txt{L}$, at the moment of sheet creation, $r_\txt{sheet}=r_\txt{L} \left( t_\txt{sheet} \right)$, at different pressures. This radius is scaled by the radius of the original drop, $r^* \equiv \frac{r_\txt{sheet}}{R}$. It is apparent from Fig.\ \ref{fig:rSheet} that the sudden change in sheet creation time occurs when  $r^* \approx 1$. At low pressures, the thin sheet is created at later times, when the liquid has spread further than the radius of the impacting drop (cf.\ top row of Fig.\ \ref{fig:interference}). At the transition pressure, $P = 58 \si{\kilo Pa}$, $r^*$ approaches unity. Upon a further increase in pressure, $r^*$ drops sharply, and at high pressures the thin sheet is created when $r^* < 1$ (cf.\ bottom row of Fig.\ \ref{fig:interference}). As the impact velocity is decreased, the transition pressure decreases as well. It is now clear why no transition between $r^* > 1$ and $r^* < 1$ was observed for $V = 1.4 \si{\m.s^{-1}}$ or $V = 3.4 \si{\m.s^{-1}}$ in Fig.\ \ref{fig:tSheet_237}. Since $r^*$ increases with decreasing pressure, the largest possible $r^*$ for a given drop will be found at the lowest possible pressure, i.e.\ $P_\txt{sheet}$. For $V = 1.4 \si{\m.s^{-1}}$, even at $P_\txt{sheet}$, $r^* < 1$. Similarly, the smallest $r^*$ will occur at the highest possible pressure, which in our case is atmospheric pressure, and for $V = 3.4 \si{\m.s^{-1}}$, $r^* \left( P_\txt{atm} \right)  > 1$.

\begin{figure}
\includegraphics[width=\columnwidth]{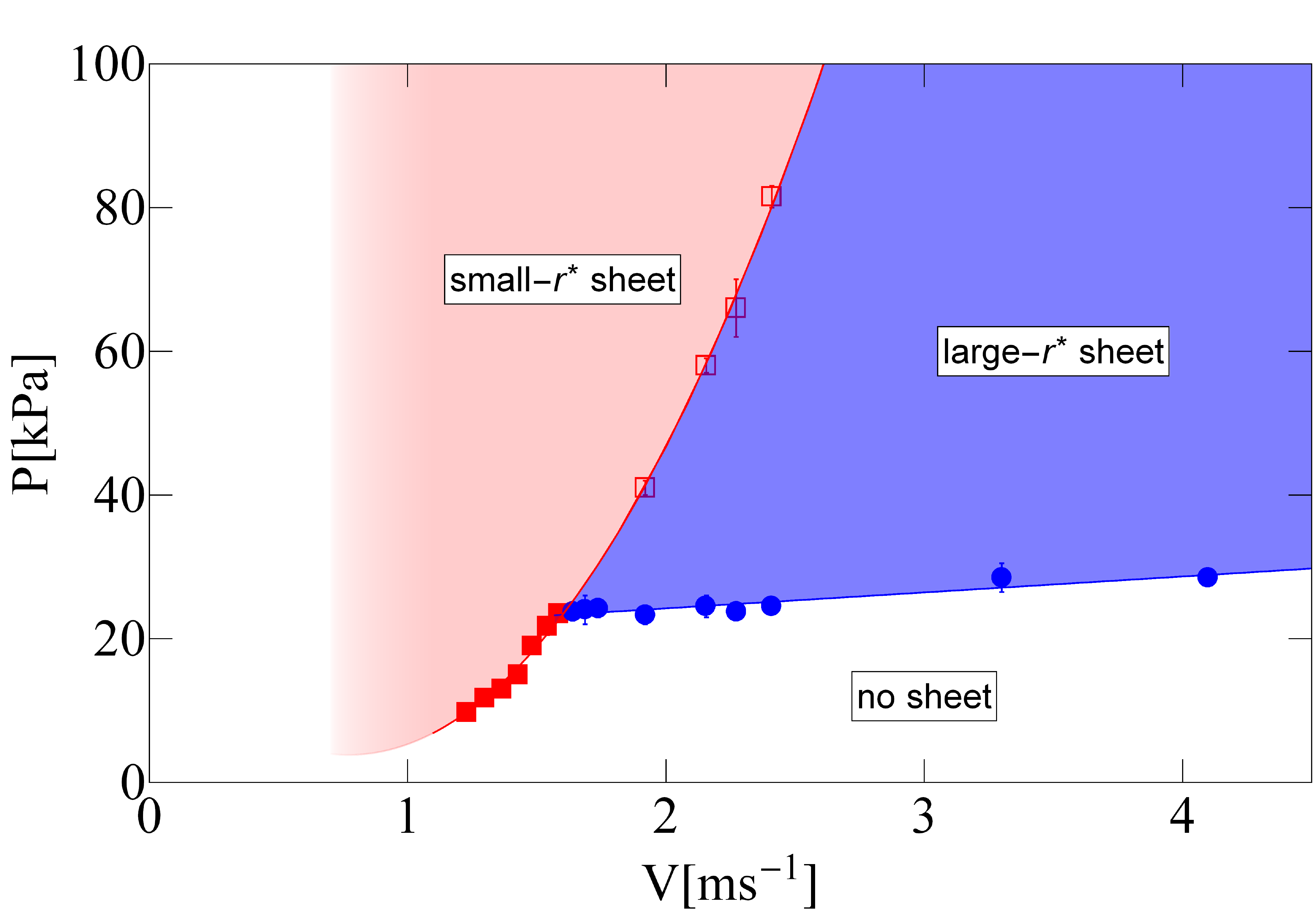}
\caption[]{\label{fig:Psheet_u0}
Diagram of drop impact outcomes for $9.4 \si{\milli Pa.s}$ silicone oil drops of radius $R=1.65 \si{\milli \m}$ as gas pressure and impact velocity are varied: small-$r^*$ sheet (red), large-$r^*$ sheet (blue), and no sheet (white). (Data was not taken below $V = 1 \si{\m.s^{-1}}$, therefore that region is left blank.) The boundaries between the regions are $P_{\txt{small-}r^*}$ (\includegraphics[height=6pt]{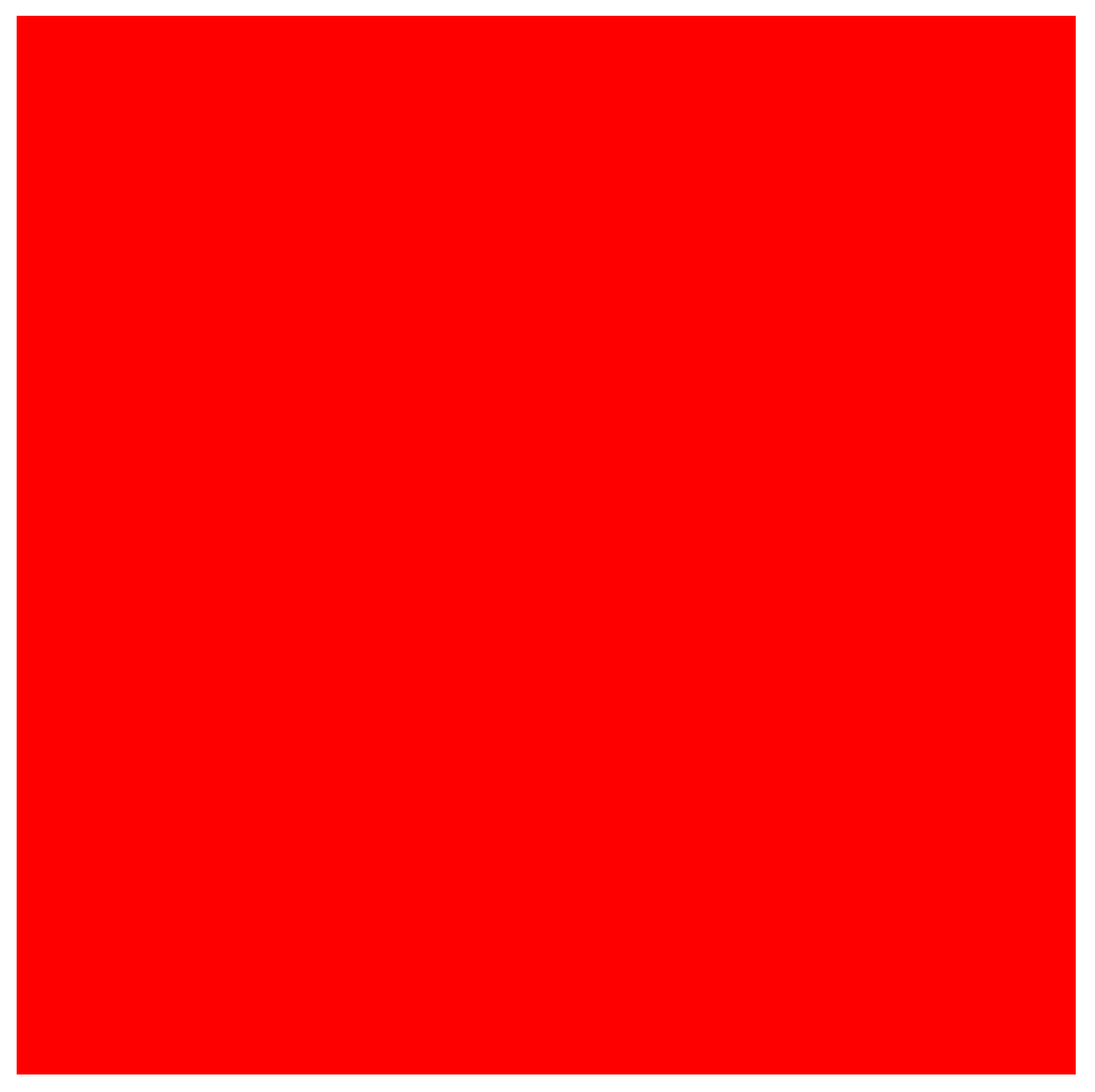}) and $P_{\txt{large-}r^*}$ (\includegraphics[height=6pt]{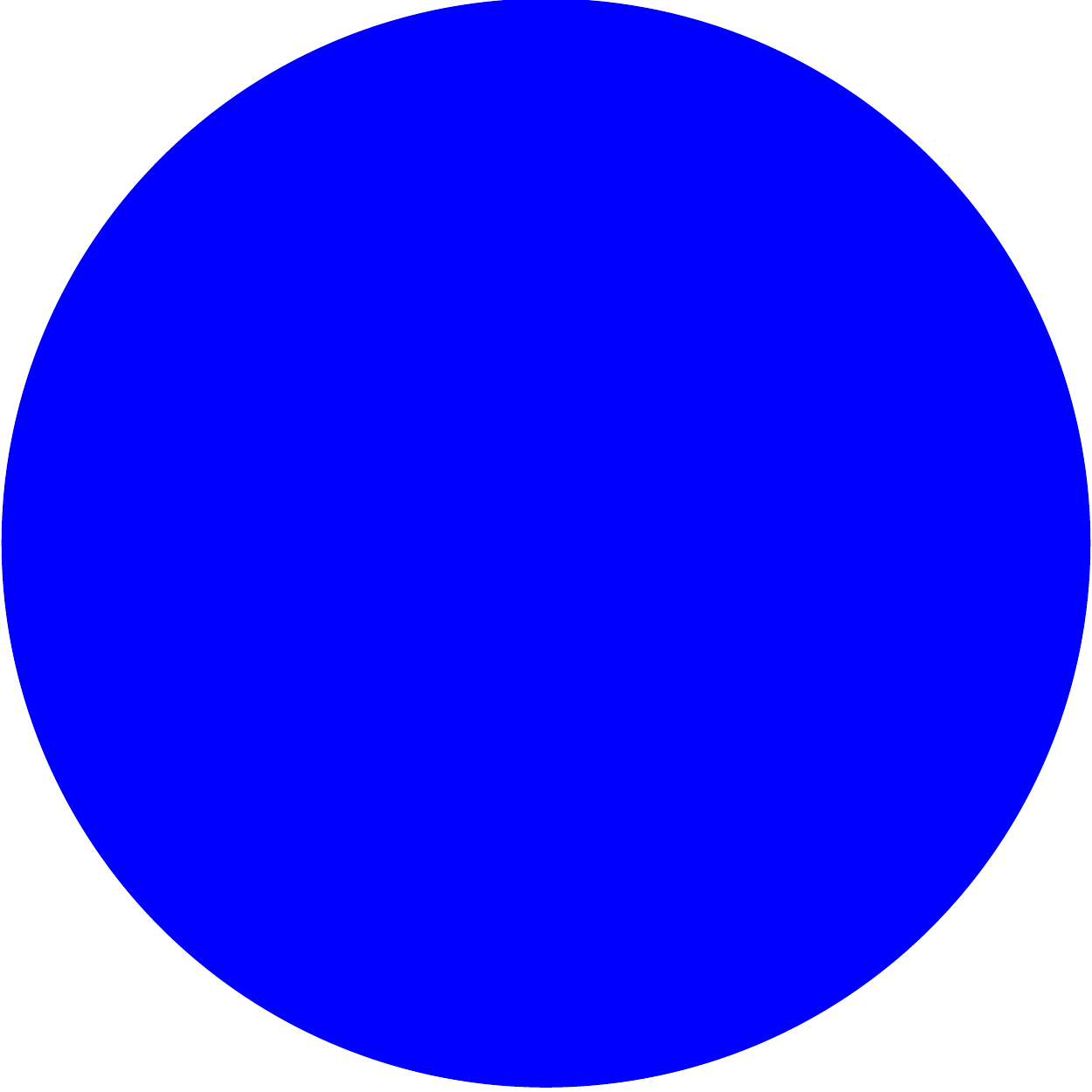}). The smaller of the thresholds at a given $V$ determines $P_\txt{sheet}$ (filled symbols). If $P_{\txt{small-}r^*} > P_{\txt{large-}r^*}$, $P_{\txt{small-}r^*}$ marks the transition between a small-$r^*$ and a large-$r^*$ sheet (empty symbols), as shown in Fig.\ \ref{fig:tSheet_237}. The lines serve as a guide to the eye. 
}
\end{figure}

\begin{figure}
\includegraphics[width=\columnwidth]{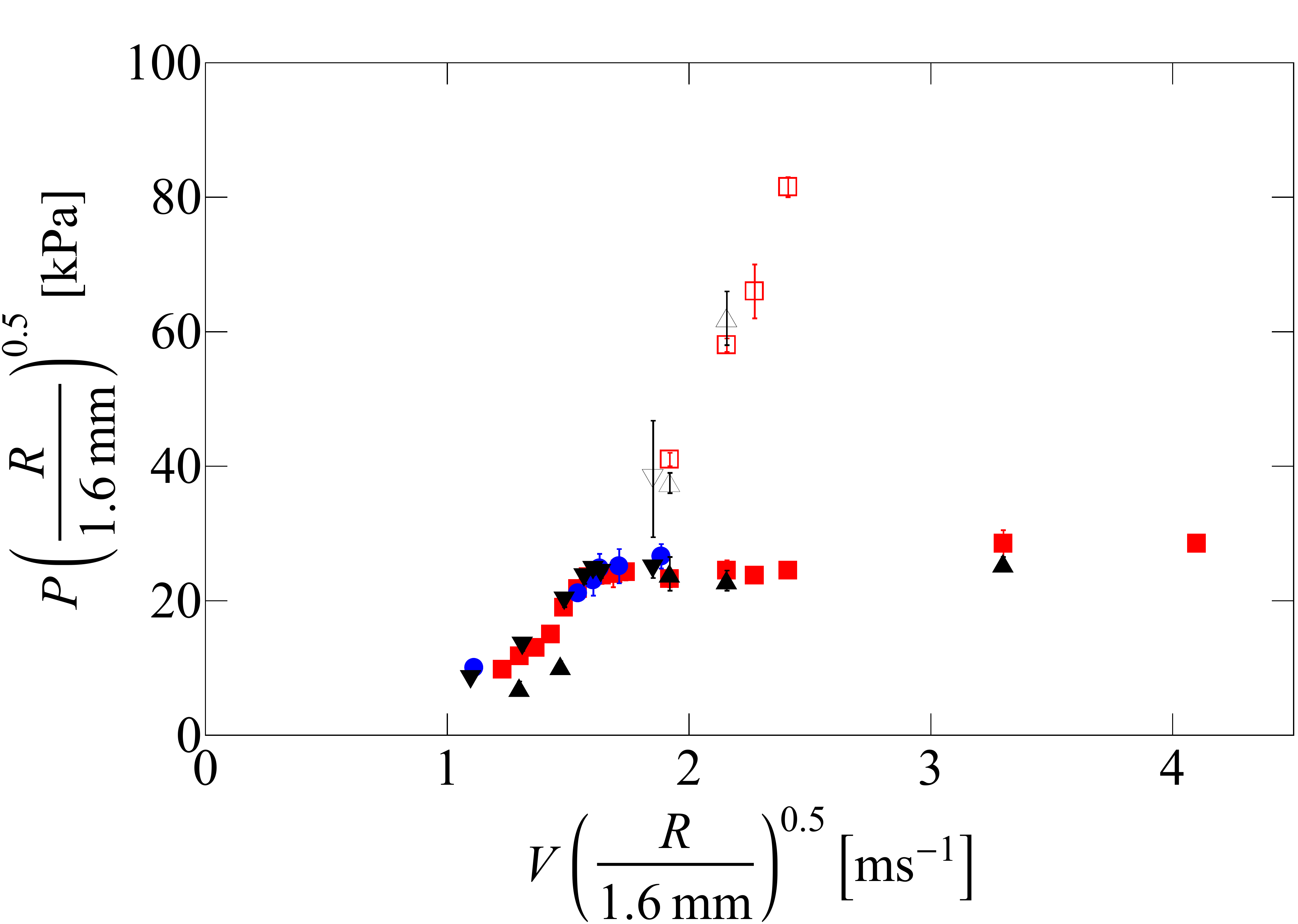}
\caption[]{\label{fig:Psheet_u0_scaled}
Threshold pressures of small-$r^*$ and large-$r^*$ sheets for silicone oil drops of viscosity $9.4 \si{\milli Pa.s}$ and radius $R = 0.85 \si{\milli \m}$ (\includegraphics[height=6pt]{FilledBlueCircle.pdf}),  $1.20 \si{\milli \m}$ (\includegraphics[height=6pt]{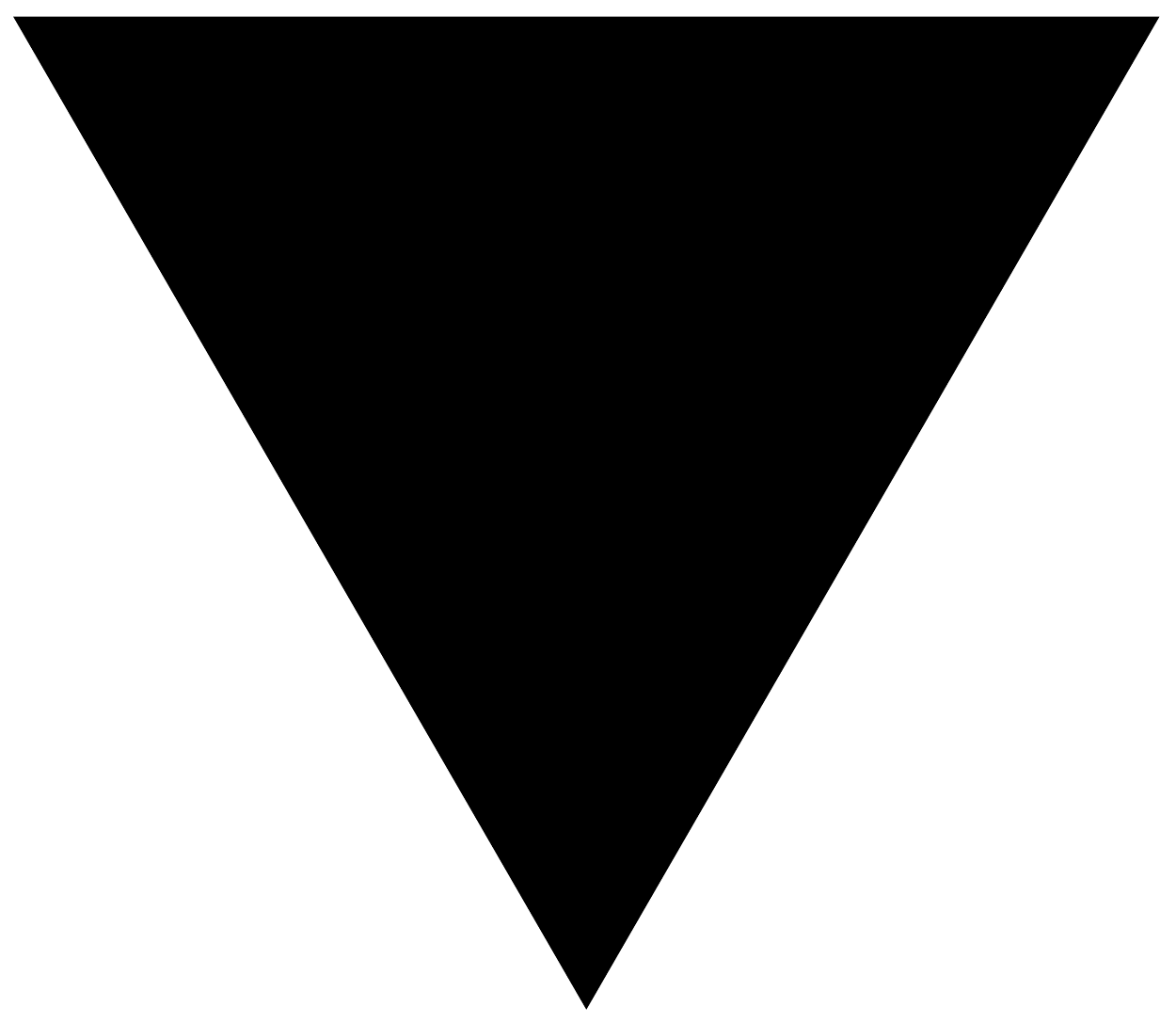}), $1.65 \si{\milli \m}$ (\includegraphics[height=6pt]{FilledRedSquare.pdf}), and $4.6 \si{\milli Pa.s}$ drops of radius $1.65 \si{\milli \m}$ (\includegraphics[height=6pt]{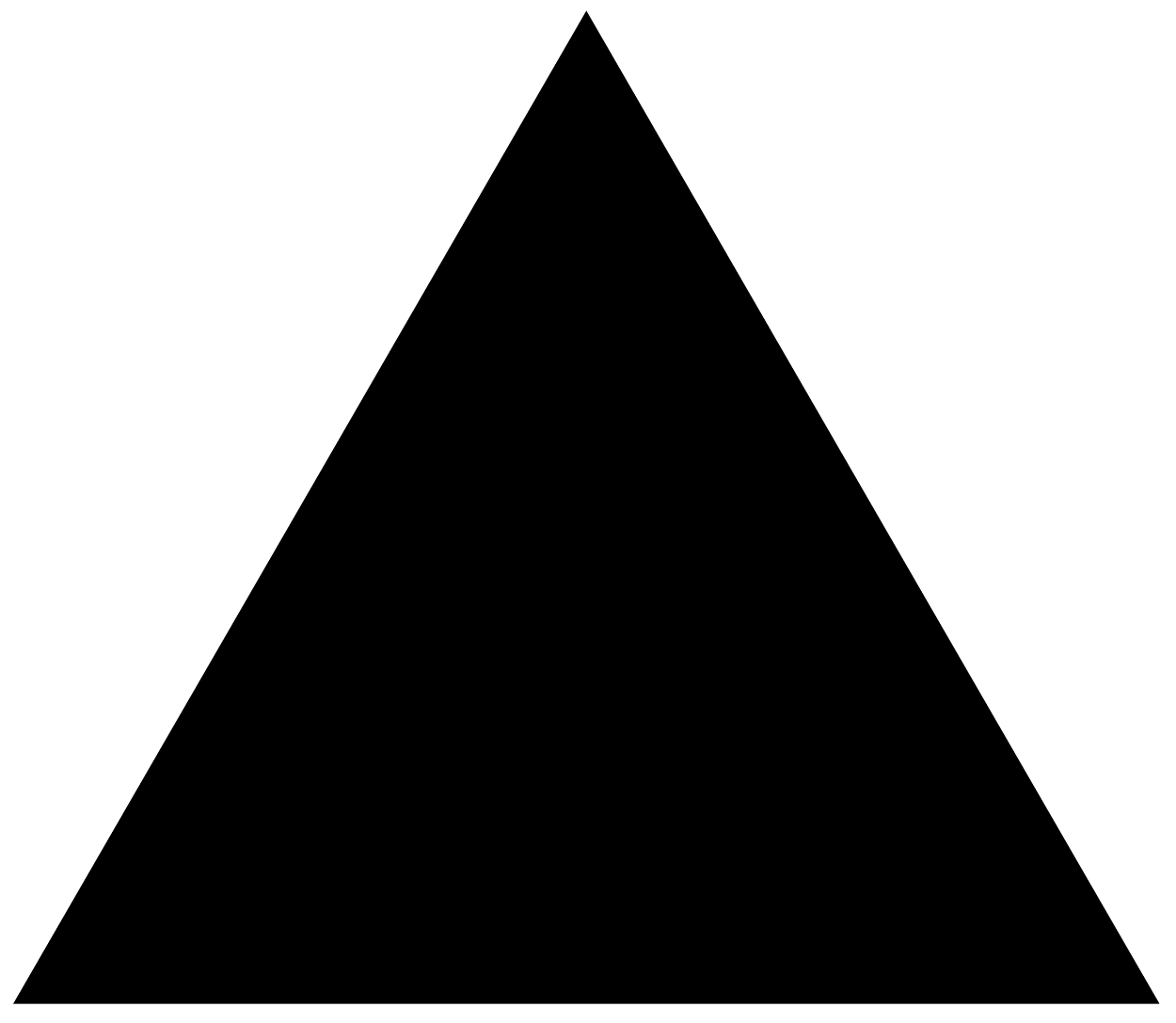}). The open symbols represent the transition between the small-$r^*$ and large-$r^*$ sheets. The data is scaled by $R^{0.5}$.
}
\end{figure}

The data in Fig.\ \ref{fig:rSheet} allows one to determine which sheet, small-$r^*$ or large-$r^*$, will be created at a given pressure and impact velocity. Figure \ref{fig:Psheet_u0} summarizes the possible outcomes of drop impact in a phase diagram. Consider, for example, a drop with $V = 2.2 \si{\m.s^{-1}}$. For pressures greater than $P = 58 \si{\kilo Pa}$, the sheet is created at $r^* < 1$: the outcome is a small-$r^*$ sheet, represented by the red region. Below $58 \si{\kilo Pa}$, the sheet obtained is large-$r^*$, for which $r^*>1$, marked by the blue region. However, if the pressure is reduced below $24.5 \si{\kilo Pa}$, no sheet will be created and the corresponding region is blank. When impact velocity is increased, the pressure separating the two regions increases as well (cf. Fig.\ \ref{fig:rSheet}). Therefore, as described above, low-$V$ drops always create a small-$r^*$ sheet, as in the case of $V = 1.4 \si{\m.s^{-1}}$ in Fig.\ \ref{fig:rSheet}, while high-$V$ drops create a large-$r^*$ sheet even at atmospheric pressure, as exemplified by the $V = 3.4 \si{\m.s^{-1}}$ impacts therein. 

The phase diagram can be understood by considering distinct thresholds for small-$r^*$ and large-$r^*$ sheets. The threshold $P_{\txt{small-}r^*}$ is defined as the pressure below which the small-$r^*$ sheet is suppressed, either due to no sheet being created (\includegraphics[height=6pt]{FilledRedSquare.pdf}), or due to a transition to a large-$r^*$ sheet (\includegraphics[height=6pt]{EmptyRedSquare.pdf}), as in Fig.\ \ref{fig:rSheet}. Similarly, the large-$r^*$ sheet is possible above $P_{\txt{large-}r^*}$ (\includegraphics[height=6pt]{FilledBlueCircle.pdf}). The two thresholds depend differently on $V$. Consequently, they cross at a point and thus yield the three regions shown in Fig.\ \ref{fig:Psheet_u0}. If $P > P_{\txt{small-}r^*}$, the thin sheet is created soon after impact, when the drop has not had a chance to spread significantly, $r^* < 1$. Note that once a sheet is created at $r^* < 1$ the drop begins to splash and a second sheet cannot be created. Therefore no large-$r^*$ sheet will be created in this case even if $P > P_{\txt{large-}r^*}$ and it is impossible to measure $P_{\txt{large-}r^*}$ for impact velocities below the crossover. If $P < P_{\txt{small-}r^*}$ the drop spreads to $r^* > 1$ and two outcomes are possible. If $P > P_{\txt{large-}r^*}$, a large-$r^*$ sheet will be created. Otherwise, if $P < P_{\txt{large-}r^*}$, the large-$r^*$ sheet will also be suppressed and no sheet creation will occur. Remarkably, the shape of the phase diagram remains unchanged for drops of different radius $R$ and viscosity $\mu$. We can approximately collapse the boundaries by scaling the impact velocity by $R^{0.5}$, as shown in Fig.~\ref{fig:Psheet_u0_scaled}.  

The existence of two types of sheet creation explains the impact velocity regimes found in \cite{Driscoll2010}. Driscoll et al.\ measured $P_\txt{sheet}$ vs.\ impact velocity and found that $P_\txt{sheet}$ depended on $V$ much more strongly at low impact velocities than at high ones. This result is reproduced here by the filled symbols of Fig.\ \ref{fig:Psheet_u0}. As defined above, the criterion for $P_\txt{sheet}$ is the presence of a thin sheet, without distinguishing whether the sheet is large-$r^*$ or small-$r^*$. Therefore, $P_\txt{sheet} = \txt{Min} \left( P_{\txt{small-}r^*},P_{\txt{large-}r^*} \right)$. In the low velocity regime, the small-$r^*$ sheet persists to lower pressures and $P_\txt{sheet}=P_{\txt{small-}r^*}$ and the dependence of sheet threshold pressure on impact velocity follows from the properties of small-$r^*$ sheet creation. Conversely, in the high-velocity regime, the sheet threshold pressure is determined by $P_{\txt{large-}r^*}$, which does not depend strongly on impact velocity.

\section{\label{sec:thresholdvelocity}Threshold velocity}

Despite extensive experimental measurements \cite{Stevens2014,Driscoll2010}, it remains a mystery why a drop creates a thin sheet at $t_\txt{sheet}$. Until now it was similarly unclear why the thin sheet never appears below the threshold pressure $P_\txt{sheet}$. The origin of this threshold is revealed by letting drops fall on surfaces with one section of the slide roughened and the other part left smooth as in in Fig.\ \ref{fig:images}.  Thin-sheet creation is suppressed on the rough region of the slide. Before $t_\txt{sheet}$, the drop looks identical on both the smooth and the rough surface. The next frame shows the emergence of a thin sheet on the smooth surface. In contrast, the drop on the partially rough surface continues to spread on the rough region and has not created a thin sheet (Fig.\ \ref{fig:images}f). In  Fig.\ \ref{fig:images}g part of the spreading liquid has reached the smooth region. There, a thin sheet begins to be created, despite the fact that $t>t_\txt{sheet}$, while the part of the drop that remains in the rough region continues to spread smoothly on the surface.  Figures \ref{fig:images}d and \ref{fig:images}h show the final stage of the splash. In both cases smaller droplets begin to break off, but in  Fig.\ \ref{fig:images}h the splash is much smaller and confined to the smooth region of the partially rough surface. Interference imaging confirms that sheet creation begins immediately when the liquid enters the smooth region.

In this fashion, it is possible to delay sheet creation. In other words, the drop has the capacity to form a sheet beginning at $t_\txt{sheet}$: before this time it will create a sheet on neither a smooth, nor a rough surface. At times $t > t_\txt{sheet}$, a sheet can be created as soon as the spreading liquid moves onto the smooth surface. However, the sheet cannot be delayed indefinitely. If the spreading liquid reaches the smooth surface at a time greater than a $t_\txt{stop}$, no thin sheet will be created. To understand the origin of $t_\txt{stop}$, we must consider the velocity of the advancing contact line.   

\begin{figure}
\includegraphics[width=1\columnwidth]{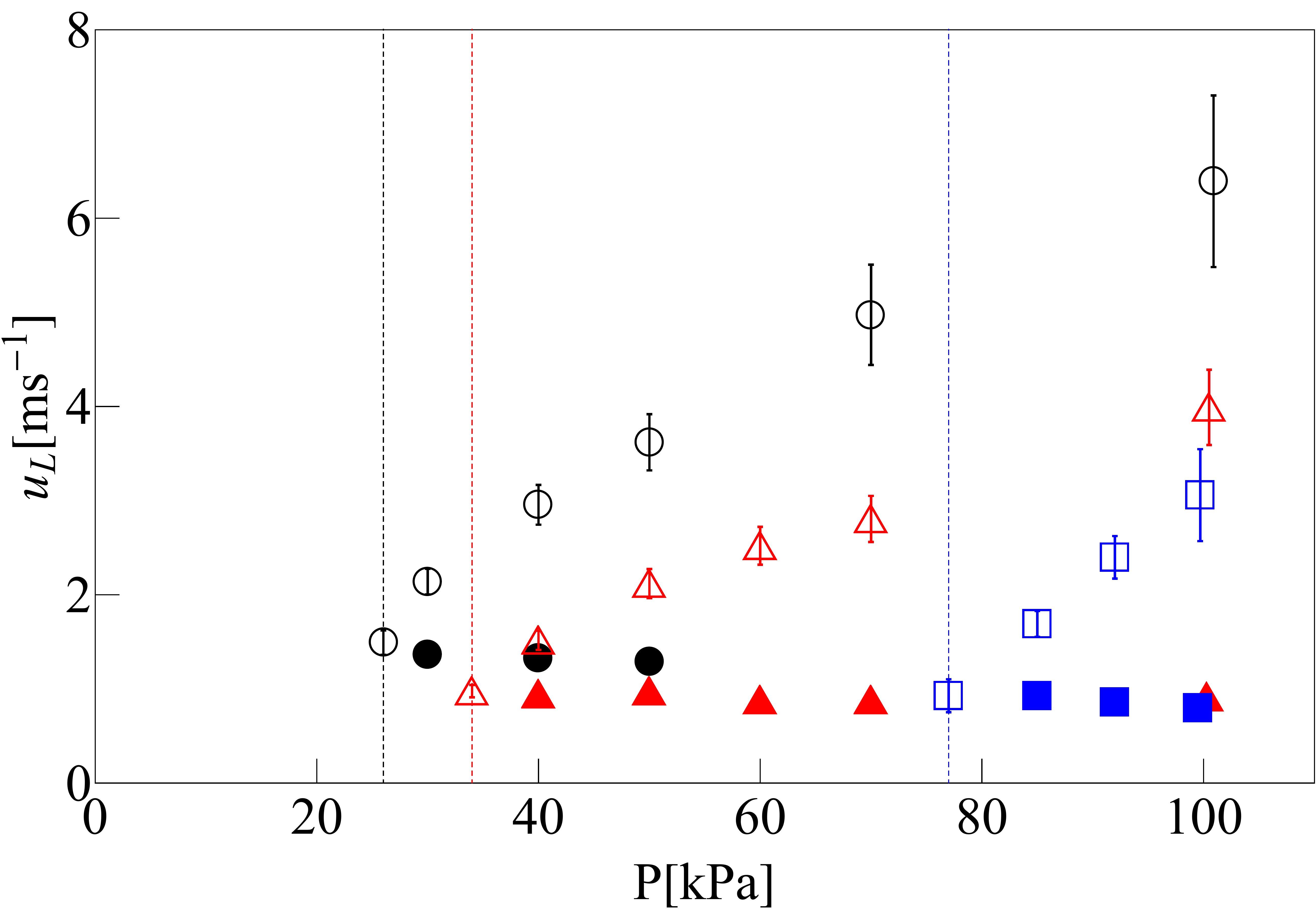}

\caption[]{\label{fig:uL_vs_pressure}
The velocity of the spreading liquid at the splashing onset $u_\txt{sheet}$ (open symbols) and the threshold velocity $u_\txt{stop}$ (closed symbols) vs.\ ambient pressure for $R = 3.2 \si{\milli m}$ silicone oil drops of viscosity $9.4 \si{\milli Pa.s}$ (\includegraphics[height=6pt]{FilledBlackCircle.pdf}), $19 \si{\milli Pa.s}$ (\includegraphics[height=6pt]{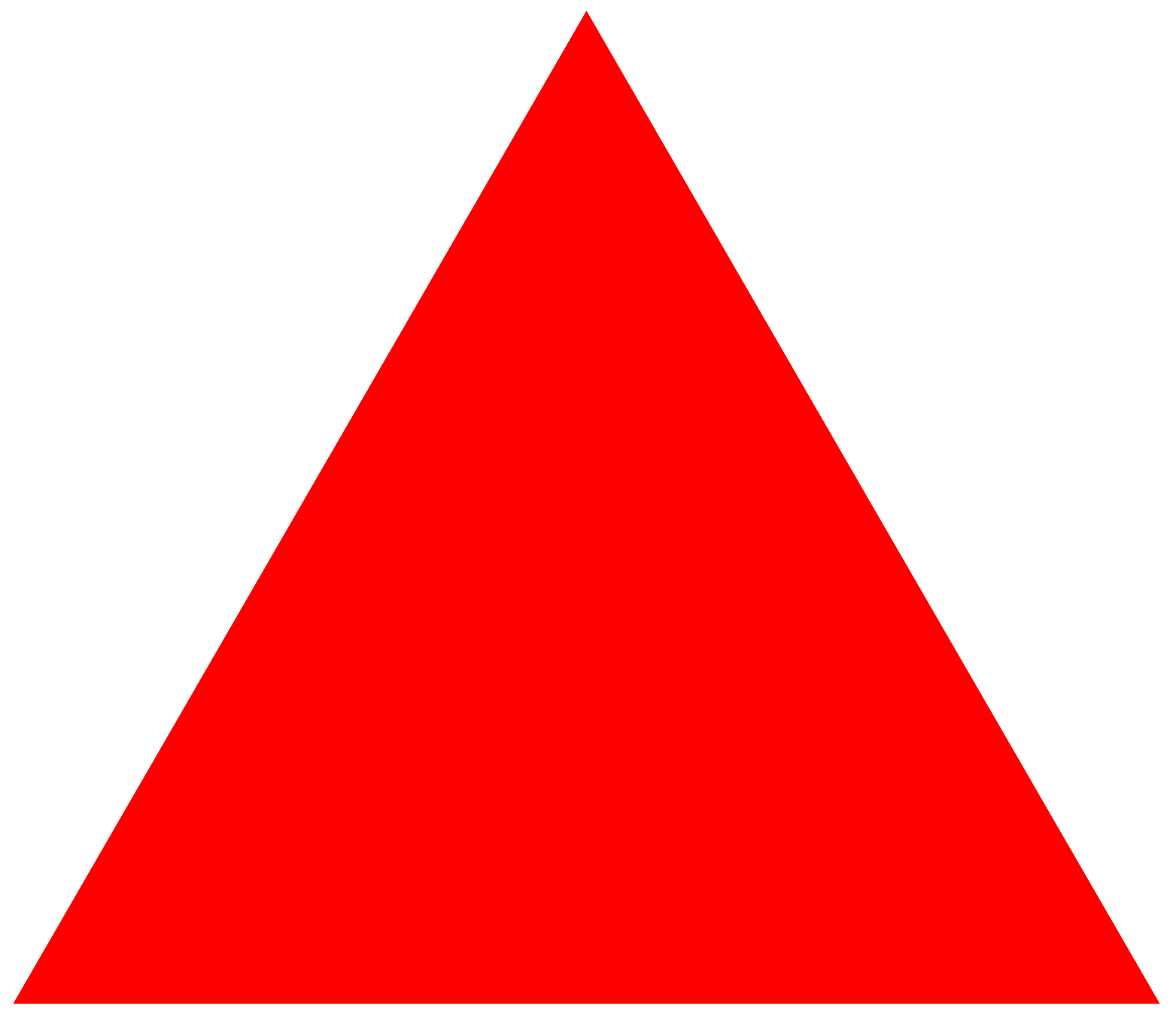}), and $48 \si{\milli Pa.s}$ (\includegraphics[height=6pt]{FilledBlueSquare.pdf}) impacting a glass slide at $3.4 \si{\m.s^{-1}}$. While $u_\txt{sheet}$ decreases with pressure, $u_\txt{stop}$ remains approximately constant. The threshold pressures $P_\txt{sheet}$, marked by dashed lines of the respective color, are set by the crossover of $u_\txt{sheet}\left( P \right)$ and $u_\txt{stop}$.
}
\end{figure}

Specifically, consider $u_\txt{sheet}$, the velocity of the contact line at the moment the thin sheet was created on the smooth surface, $u_\txt{L} \left( t_\txt{sheet} \right)$. Since as the drop spreads it is continually decelerating, a sheet created shortly after impact will have a larger $u_\txt{sheet}$ than a sheet created at a later time. This is shown by the open symbols in Fig.\ \ref{fig:uL_vs_pressure}. For example, at atmospheric pressure and on a smooth surface, the $19 \si{\milli Pa.s}$ drop (red symbols) creates a thin sheet at $t_\txt{sheet}= 0.27 \si{\milli s}$, when the liquid is spreading outward at $u_\txt{sheet} = 4.0 \pm 0.4 \si{\m.s^{-1}}$. If the sheet is delayed by having the drop fall on a partially rough surface, it will be created when the contact line is moving with a velocity smaller than $4.0 \si{m.s^{-1}}$. As the distance from the point of impact to the boundary of the smooth region increases, the velocity at which the liquid first has a chance to create a thin sheet decreases. Finally, if the velocity of the contact line is smaller than a velocity $u_\txt{stop}$ when it enters the smooth region, the thin sheet will not be created at all. Figure \ref{fig:uL_vs_pressure} shows that at atmospheric pressure and on a smooth surface, the $19 \si{\milli Pa.s}$ can not create a thin sheet if the contact line is moving slower than $u_\txt{stop}= 0.91 \pm 0.06 \si{\m.s^{-1}}$. The drop can create a sheet only when the liquid velocity $u_\txt{L}$ satisfies both conditions: $u_\txt{stop} < u_\txt{L} < u_\txt{sheet}$.

Figure \ref{fig:uL_vs_pressure} shows that as pressure decreases the splash is delayed and, consequently, $u_\txt{sheet}$ decreases, (cf. Fig.\ \ref{fig:tSheet_237}, while $u_\txt{stop}$ remains constant. Consequently there will be a pressure at which $u_\txt{sheet} \left( P \right) = u_\txt{stop}$. Below this pressure, sheet creation is impossible even on a smooth surface, since the two necessary conditions, that the sheet be created after time $t_\txt{sheet}$ and that at the moment of sheet creation the contact line be moving faster than $u_\txt{stop}$, cannot be simultaneously satisfied. Indeed, this crossover coincides with $P_\txt{sheet}$, which is marked by the dashed lines in Fig.\ \ref{fig:uL_vs_pressure}, demonstrating that $u_\txt{stop}$ determines the threshold pressure.

The  $u_\txt{stop}$ pressure independence has an important practical consequence, as it allows one to measure $u_\txt{stop}$ by simply measuring $u_\txt{sheet}\left(P_\txt{sheet}\right)$ on a smooth surface. The lowest $u_\txt{sheet}$ that can be observed is equal to $u_\txt{stop}$. This method of measuring $u_\txt{stop}$ is much more straightforward, as it does not require the use of rough surfaces. Moreover it is effective even for drops of low viscosity, whose thin sheet cannot be easily suppressed with roughness \cite{Latka2012}.


\begin{figure}
\includegraphics[width=1\columnwidth]{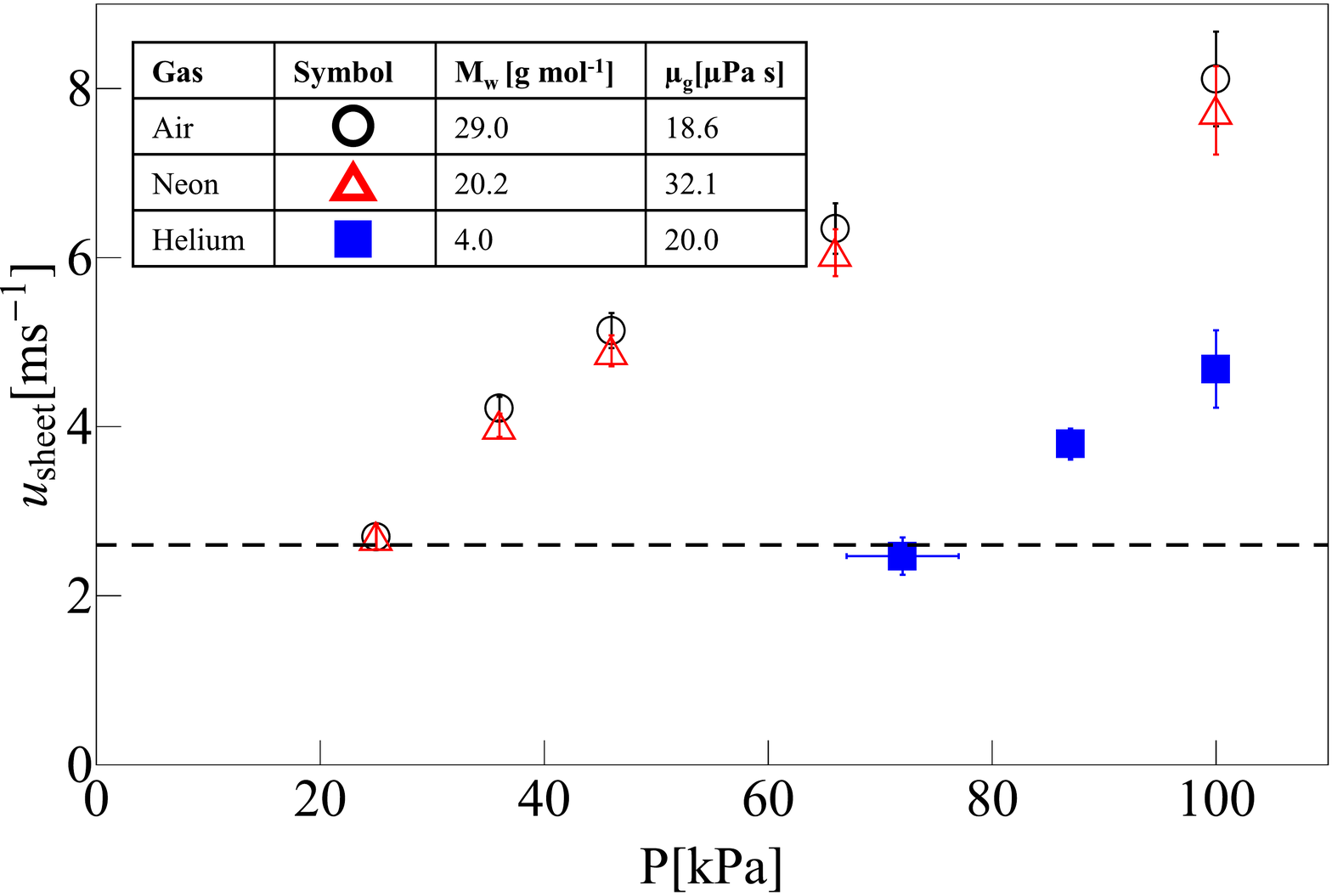}

\caption[]{\label{fig:gas_properties}
The velocity of the advancing liquid at the splashing onset $u_\txt{sheet}$ vs. gas pressure for $D=3.3 \si{\milli m}$ silicone oil drops of viscosity $4.6 \si{\milli Pa.s}$ impacting a smooth glass slide at $3.5 \si{\m.s^{-1}}$ in an atmosphere of air, neon, and helium. The dashed line marks $u_\txt{stop}=2.6 \si{\m.s^{-1}}$. For each gas, the leftmost points represent the measured $P_\txt{sheet}$.
}
\end{figure}

The pressure independence of $u_\txt{stop}$ is remarkable, given the dramatic influence the ambient gas has on thin-sheet creation. To further investigate the effect of the air, I varied the gas viscosity $\mu_g$  and molecular weight $M_w$, all of which were found to affect $t_\txt{sheet}$ \cite{Driscoll2010, Stevens2015}. The result is shown in Fig.\ \ref{fig:gas_properties} for impacts of $4.6 \si{\milli Pa.s}$ silicone oil drops. Air ($M_w = 29.0 \si{\g.\mol^{-1}}$, $\mu_g=18.6 \si{\micro Pa.s}$) and neon ($M_w = 20.2 \si{\g.\mol^{-1}}$, $\mu_g=32.1 \si{\micro Pa.s}$) have comparable $M_w$ and $u_\txt{sheet}$ at each pressure. As expected, helium ($M_w = 4.0 \si{\g.\mol^{-1}}$, $\mu_g=20.0 \si{\micro Pa.s}$) has a lower $u_\txt{sheet}$ than neon and air, due to its much lower $M_w$  \cite{Driscoll2010}, however the threshold velocity $u_\txt{stop}$ is the same for all gases. Therefore, the threshold pressure can be understood as the combination of an air-dependent mechanism that creates a thin sheet, and $u_\txt{stop}$, an air-independent threshold of stability below which this mechanism is not possible. An example of this is the higher $P_\txt{sheet}$ in helium compared to air or neon. Since at a given pressure sheet creation occurs later for gases with lower $M_w$, the threshold pressure at which $u_\txt{sheet}$ intersects $u_\txt{stop}$ will be higher, as shown in Fig.\ \ref{fig:gas_properties}.  

\begin{figure}
\includegraphics[width=\columnwidth]{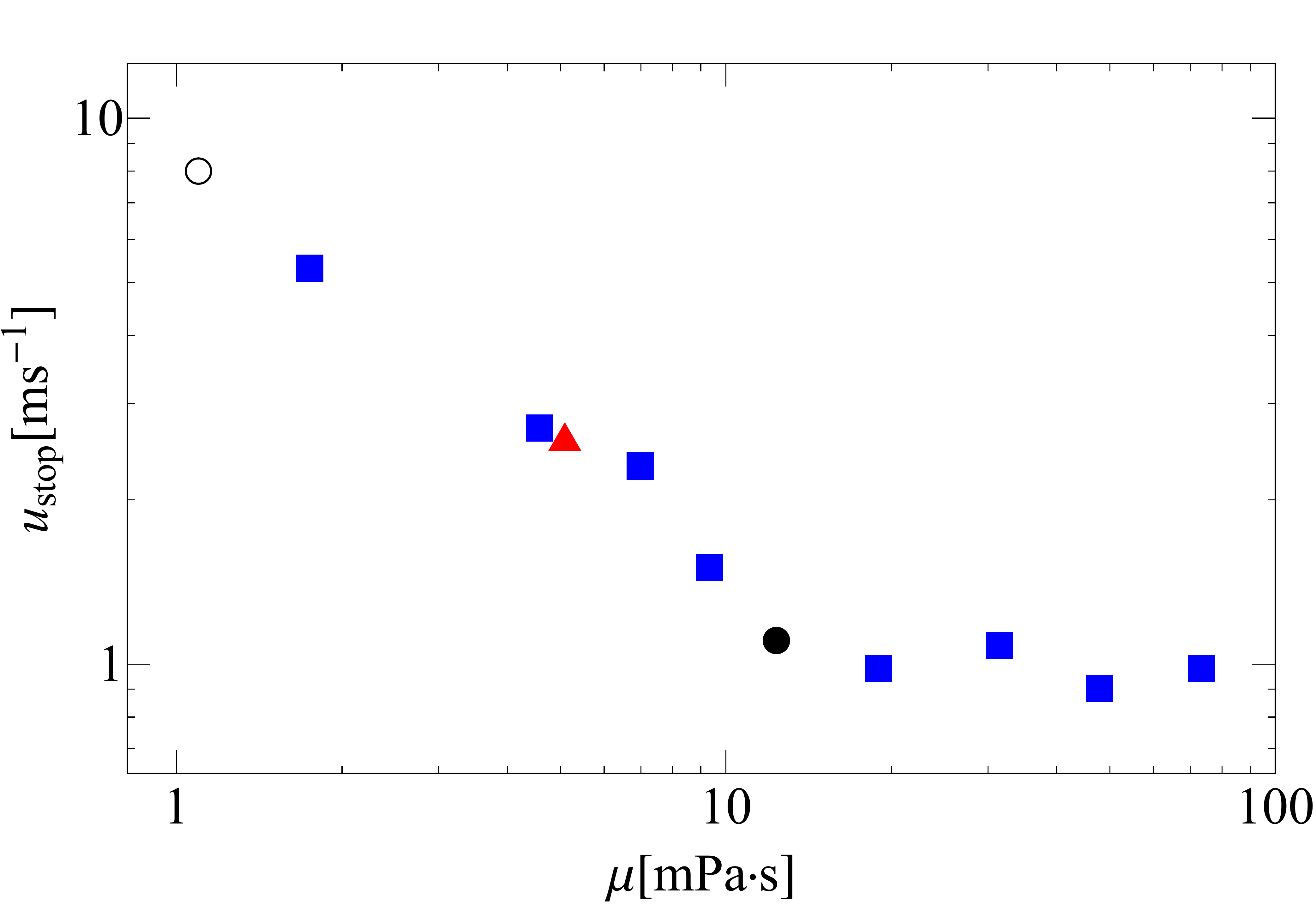}

\caption[]{\label{fig:uStop_liquid_properties}
A log-log plot of the threshold velocity $u_\txt{stop}$ vs. liquid viscosity $\mu_\txt{L}$ for $D=3.3 \si{\milli \m}$ drops impacting a glass slide at $3.4 \si{\m.s}$ in an atmosphere of air. Various liquids were used: ethanol (\includegraphics[height=6pt]{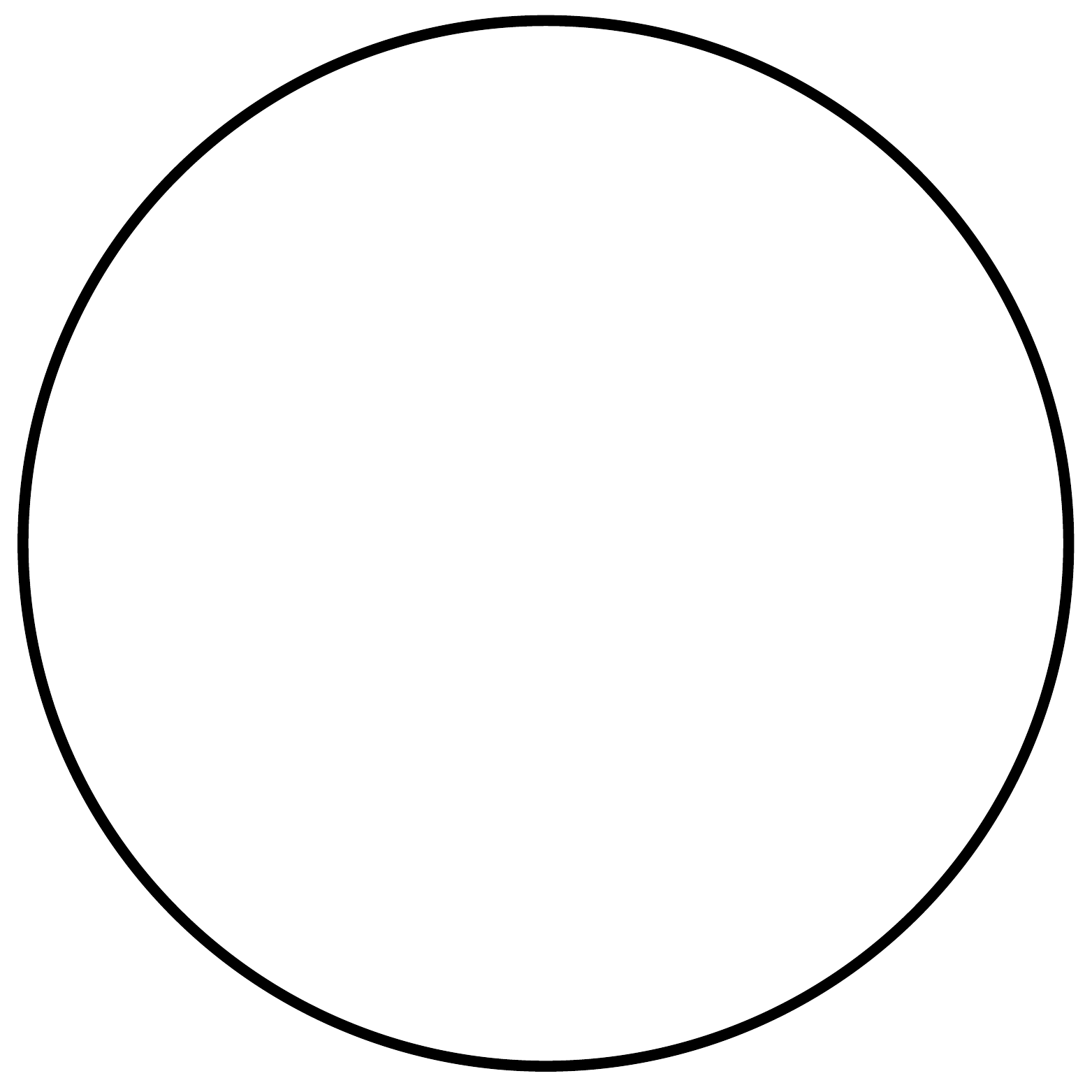}), silicone oil (\includegraphics[height=6pt]{FilledBlueSquare.pdf}), a solution of ethanol, water and sucrose (\includegraphics[height=6pt]{FilledRedTriangle.pdf}), and a solution of water and glycerol (\includegraphics[height=6pt]{FilledBlackCircle.pdf}).
}
\end{figure}

\begin{figure}
\includegraphics[width=\columnwidth]{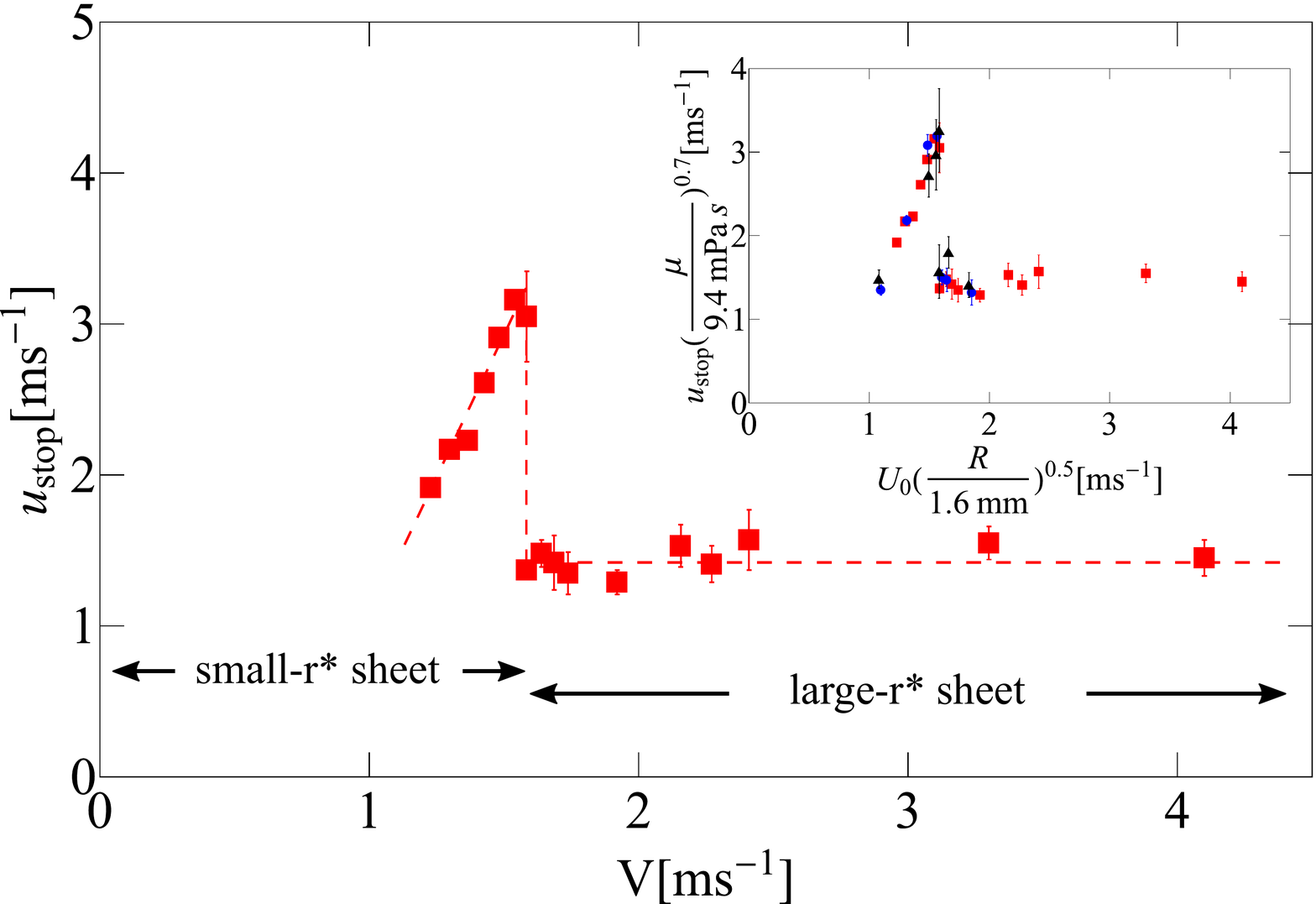}
\caption[]{\label{fig:uStop_velocity}
Threshold velocity $u_\txt{stop}$ vs. impact velocity for  for $9.4 \si{\milli Pa.s}$ silicone oil drops of radius $1.6 \si{\milli \m}$ (\includegraphics[height=6pt]{FilledRedSquare.pdf}). A discontinuous transition separates a regime where $u_\txt{stop}$ is dependent on a low impact velocity and one where $u_\txt{stop}$ is independent of high impact velocities. The two regimes correspond to the small-$r^*$ and large-$r^*$ sheets. The dashed lines are a guide to the eye. Inset:  $u_\txt{stop}$ vs. $V$ scaled by $\left( \dfrac{R}{1.6 \si{\milli \m}} \right)^{0.5}$ for $9.4 \si{\milli Pa.s}$ silicone oil drops of radius $0.8 \si{\milli \m}$ (\includegraphics[height=6pt]{FilledBlackUpTriangle.pdf}), $1.2 \si{\milli \m}$ (\includegraphics[height=6pt]{FilledBlueCircle.pdf}), and $1.6 \si{\milli \m}$ and for a $4.6 \si{\milli Pa.s}$, $1.6 \si{\milli \m}$ drop (\includegraphics[height=6pt]{EmptyRedSquare.pdf}) The data is collapsed by scaling with the drop radius and liquid viscosity.
}
\end{figure}

The threshold velocity $u_\txt{stop}$ does show a significant dependence on the liquid viscosity. Figure \ref{fig:uStop_liquid_properties} shows that $u_\txt{stop}$ decreases approximately as $\mu^{-0.7 \pm 0.1}$ for liquid viscosities below $20 \si{\milli Pa.s}$. Changing the surface tension does not change $u_\txt{stop}$. In Fig.\ \ref{fig:uStop_liquid_properties}, I increase the surface tension by over a factor of two with respect to silicone oil, by using a mixture either of ethanol, water and sucrose ($\mu=5.2 \si{\milli Pa.s}$, $\sigma=55 \si{\milli N.m}$) or water and glycerol ($\mu=12.3 \si{\milli Pa.s}$, $\sigma=66 \si{\milli N.m}$). In both cases, the measured $u_\txt{stop}$ is similar to that of a silicone oil drop. Similarly, changing drop size, drop density and the wetting properties \cite{latka2016drop} of the surface failed to change $u_\txt{stop}$. 

The dependence of $u_\txt{stop}$ on impact velocity $V$, shown in Fig.\ \ref{fig:uStop_velocity}, shows two distinct regimes. At large impact velocities, $u_\txt{stop}$ is independent of impact velocity and equal to the value that corresponds to Fig.\ \ref{fig:uStop_liquid_properties}. However, if the impact velocity is decreased, one observes a discontinuous transition to a regime in which $u_\txt{stop}$ is approximately linear with $V$. This transition corresponds to the crossover seen in Fig.\ \ref{fig:Psheet_u0}.  Below the crossover, $u_\txt{stop}$ is determined by the properties of the small-$r^*$ sheet, reaching up to $3.16 \pm 0.05 \si{m.s^{-1}}$ at $V = 1.54 \si{m.s^{-1}}$. Above the crossover, a large-$r^*$ is observed even after the small-$r^*$ has been suppressed. Therefore, $u_\txt{stop}$ is set by the large-$r^*$ sheet and is reduced to $1.37 \pm 0.06 \si{m.s^{-1}}$ at $V = 1.58 \si{m.s^{-1}}$. The discontinuous change in the threshold velocity underscores the difference between small-$r^*$ and large-$r^*$ sheets. Note that results presented in Figs.\ \ref{fig:uL_vs_pressure}-\ref{fig:uStop_liquid_properties} are all measured for large-$r^*$ sheets.

The inset of Fig.\ \ref{fig:uStop_velocity} shows $u_\txt{stop}$ data for drops of varying radius and liquid viscosity. Since the discontinuous transition results from the change in the type of thin sheet, it is to be expected that its position will scale in the same way as the crossover between $P_{\txt{small-}r^*}$ and $P_{\txt{large-}r^*}$ in Fig.\ \ref{fig:Psheet_u0_scaled}. At the same time, the effect of viscosity on $u_\txt{stop}$ should follow the scaling found in Fig.\ \ref{fig:uStop_liquid_properties}. Indeed, the data is collapsed by scaling the x-axis by $R^{0.5}$ and the y-axis by $\mu^{0.7}$.

\section{\label{sec:conclusion}Discussion and conclusions}

The cause of the sharp transition between small-$r^*$ and large-$r^*$ sheets can potentially be understood by considering the geometry of the drop. Early during impact, the drop has not spread far along the substrate, as shown in the top image of Fig.\ \ref{fig:rSheet}. The spreading liquid is separated from the main drop by a region of sharp curvature where the undeformed drop meets the spreading liquid. In contrast, when  $r^* > 1$ this curvature is gone. It is reasonable to believe that the flows within the drop are significantly different in the two cases and that the difference in liquid flow causes the change in dependence on pressure in the two types of sheets.

Regardless of the underlying physical mechanism, one must be careful in comparing results obtained in different impact velocity regimes. First, the high and low impact velocity regimes have been distinguished by the velocity $V^*$ at which $P_\txt{sheet}$ changes its dependence on $V$ \cite{Stevens2014}. Figure \ref{fig:Psheet_u0} shows that the relevant boundary between small-$r^*$ and large-$r^*$ sheets varies with pressure. Consequently, one cannot specify impact velocity regimes without additionally specifying the gas pressure. Second, it follows from Fig.\ \ref{fig:Psheet_u0} that if one is close to the boundary between small-$r^*$ and large-$r^*$ sheets, a small increase in $V$ could lead to a large increase in $t_\txt{sheet}$. Since a sheet that is created at a larger $r^*$ leads to a smaller splash, an increase in impact velocity can lead to a seemingly paradoxical decrease in the size of the resulting splash. Additional work is needed to reconcile theoretical models, simulations, and experiments that did not account for this effect. 

The swift spreading of the liquid-air-solid contact line during a splash has provoked numerous comparisons with forced wetting or coating experiments \cite{Rein2008,Thoroddsen2010,Driscoll2010}. In the simplest example of forced wetting, a solid is plunged vertically into a liquid bath with a given velocity. If the solid enters the bath slower than a critical velocity $u_\txt{critical}$, the contact line remains stable. If the contact line velocity is forced to exceed $u_\txt{critical}$, then the contact line becomes unstable and entrains air \cite{Perry1967}. A liquid drop spreading after impact represents a remarkably similar process. Here as well the contact line is forced to move with a given velocity. At $t_\txt{sheet}$ the contact line ceases to be stable so that an air gap appears, as long as the contact line is moving above the threshold $u_\txt{stop}$.

The similarities between $u_\txt{critical}$ in the forced wetting problem and $u_\txt{stop}$ are not merely qualitative. Gutoff and Kendrick, for example,  find that $u_\txt{critical}$ scales as $\mu^{-0.67}$ \cite{Gutoff1982}, compared to $\mu^{-0.7 \pm 0.1}$ as found here. Furthermore, forced wetting has in general been found to be independent of the ambient gas pressure, except at pressures an order of magnitude smaller than the ones considered here \cite{Benkreira2008}. Therefore, at comparable pressures, both $u_\txt{stop}$ and $u_\txt{critical}$ are gas independent.

Notably, $u_\txt{critical}$ and $u_\txt{stop}$ differ with respect to surface tension. The majority of discussions of forced wetting successfully describe $u_\txt{critical}$ in terms of the capillary number $\txt{Ca}=\dfrac{\mu V}{\sigma}$, which compares the viscosity and surface tension forces across an interface. In contrast, the threshold velocity in splashing shows a clear dependence only on the liquid viscosity: $u_\txt{stop}=f\left(\mu^{-0.7}, ... \right)$. The remaining parameters, from which $D$, $V$, $\rho$, and $\sigma$ have now  been excluded by the experiments reported above, must still yield units of $\si{\m.s^{-1}}$ on the left-hand side of the equation. Nevertheless, some experiments fail to show a surface tension dependence of $u_\txt{critical}$ in forced wetting and thus the role of surface tension in forced wetting is usually taken to be secondary to that of viscosity \cite{Benkreira2008}. Further research in both forced wetting and splashing is necessary before the role of surface tension in both processes can be understood.

Existing experiments \cite{Kolinski2012} and a recent simulation \cite{boelens2016observation} confirm that early during drop impact the contact line does in fact behave differently in the early and late stages of spreading. Kolinski et al.\ \cite{Kolinski2012} have found that after the drop impacts the solid surface and the liquid edge is forced to move rapidly $\left( \txt{Ca} \gg 1 \right)$, the liquid edge does not immediately make contact with the surface. Instead it traps an ultra-thin layer of air that persists for only a few microseconds. This is unlike the thicker air gap that causes thin-sheet creation, before collapsing due to van der Waals forces \cite{Kolinski2012}. This type of contact line motion is qualitatively different from what has been observed for contact lines driven slowly $\left( \txt{Ca} < 1 \right)$. A possible explanation of $u_\txt{stop}$ is that the unknown mechanism by which air causes thin-sheet creation requires the contact line motion described by Kolinski et al. In this case, $u_\txt{stop}$ would be the velocity above which this motion is possible. 

The process by which a drop impacting a solid substrate forms a splash involves the intermediate step of creating a thin sheet of liquid that subsequently breaks apart into secondary droplets. The properties of the ambient air, primarily the pressure, determine the time after impact when this occurs.  The present work focuses on two crucial aspects of sheet creation. First, I find a stark contrast between small-$r^*$ sheets created shortly after impact and large-$r^*$ sheets created at later times, when the flows inside the drop have qualitatively changed. This distinction accounts for the presence of a high and a low impact velocity regime, and is vital to a proper interpretation of experimental results. Second, a vast majority of theoretical and experimental work on splashing has focused on measuring the threshold pressure, below which splashing is suppressed, without explicitly considering thin-sheet creation. Here, I showed that the threshold pressure is set by two distinct conditions: the air-dependent time of thin-sheet creation, and the air-independent threshold velocity $u_\txt{stop}$ that is related to contact line stability. Together, these results emphasize that a successful theory for splashing must focus on the creation of the thin sheet.
\begin{acknowledgments}

I am deeply grateful to Sidney R. Nagel for his guidance and mentoring. This work was supported by the University of Chicago Materials Research and Engineering Center (MRSEC) through grant DMR-1420709 and by the NSF grant DMR-1404841. 

\end{acknowledgments}

\bibliographystyle{apsrev4-1}
\bibliography{uStop.bib}
\end{document}